\documentclass[conference, 10pt]{IEEEtran}

\usepackage{cite}
\usepackage{url}
\usepackage{graphicx}
\usepackage{color}
\usepackage{placeins}
\usepackage{float}
\usepackage{tabularx,colortbl}
\usepackage{ifthen}
\usepackage{gensymb}
\usepackage{textcomp}
\usepackage{gensymb}
\usepackage{comment}
\usepackage{amsmath,amssymb,amsthm}
\usepackage{verbatim}
\usepackage{subcaption}
\usepackage{multirow}
\makeatletter

\newcounter{author}
\renewcommand{\author}[2][]{
   \stepcounter{author}
   \@namedef{author@\theauthor}{#2}
   \@namedef{authorlabel@\theauthor}{#1}
}

\newcounter{address}
\newcommand{\address}[2][]{
   \stepcounter{address}
   \@namedef{address@\theaddress}{#2}
   \@namedef{addresslabel@\theaddress}{#1}
}

\newcommand{\alsep}{and}

\def\newmaketitle{\par%
  \begingroup%
  \normalfont%
  \def\thefootnote{}%  the \thanks{} mark type is empty
  \def\footnotemark{}% and kill space from \thanks within author
  \let\@makefnmark\relax% V1.7, must *really* kill footnotemark to remove all \textsuperscript spacing as well.
  \footnotesize%       equal spacing between thanks lines
  \footnotesep 0.7\baselineskip%see global setting of \footnotesep for more info
  \normalsize%
  \twocolumn[\thenewmaketitle\@IEEEaftertitletext]%
  \if@IEEEusingpubid
     \enlargethispage{-\@IEEEpubidpullup}%
  \fi
  \endgroup
  \setcounter{footnote}{0}\let\maketitle\relax\let\@maketitle\relax
  \gdef\@thanks{}%
  \let\thanks\relax}

\def\thenewmaketitle{
  \newpage
  \begin{center}%
    \vskip0.2em{\Huge\@IEEEcompsoconly{\sffamily}\@IEEEcompsocconfonly{\normalfont\normalsize\vskip 2\@IEEEnormalsizeunitybaselineskip
   \bfseries\large}\@title\par}\vskip1.0em\par%
    \vspace{1ex}
    \newcounter{c@author}
    \newcounter{c@tmp}
    \ifthenelse{\value{author}=2}{%
      \newcommand{\liand}{ and }}{%
      \newcommand{\liand}{, and }}
    \ifthenelse{\value{address}<2}{%
      \@nameuse{author@1}%
      \stepcounter{c@author}%
      \whiledo{\value{c@author}<\value{author}}{%
        \setcounter{c@tmp}{\value{author}}%
        \addtocounter{c@tmp}{-\value{c@author}}%
        \ifthenelse{\value{c@tmp}=1}{%
          \renewcommand{\alsep}{\liand}}{\renewcommand{\alsep}{, }}%
        \stepcounter{c@author}\alsep \@nameuse{author@\thec@author}}\\%
    }
    {%Add address references after the author's name
      \@nameuse{author@1}${}^{(\ref{\@nameuse{authorlabel@1}})}$%
      \stepcounter{c@author}%
      \whiledo{\value{c@author}<\value{author}}{%
      \setcounter{c@tmp}{\value{author}}%
      \addtocounter{c@tmp}{-\value{c@author}}%
      \ifthenelse{\value{c@tmp}=1}{%
        \renewcommand{\alsep}{\liand}}{\renewcommand{\alsep}{, }}%
      \stepcounter{c@author}\alsep \@nameuse{author@\thec@author}%
        ${}^{(\ref{\@nameuse{authorlabel@\thec@author}})}$%
      }
    }
    \vspace{0.2ex}

    \ifthenelse{\value{address}>0}{%
      \ifthenelse{\value{address}=1}{
        {\@nameuse{address@1}}
      }
      {%Output the addresses as an enumerated list
        \newcounter{c@address}

        \begin{center}
        \whiledo{\value{c@address}<\value{address}}
        {
          \refstepcounter{c@address}
            ${}^{(\thec@address)}$\,%
              \label{\@nameuse{addresslabel@\thec@address}}%
              \@nameuse{address@\thec@address}\\ %
        }
        \end{center}
      } % end of the address creation ifthenelse block
    }
    {
      \relax
    }
  \end{center}
}

\makeatother

\title{Experimental Analysis of Biasing Voltage Generation in Wave-Controlled RIS}

\author[org1]{Miguel Saavedra-Melo}
\author[org1]{Benjamin Bradshaw}
\author[org1]{Vanessa Yao}
\author[org1]{Ender Ayanoglu}
\author[org1]{Lee Swindlehurst}
\author[org1]{Filippo Capolino}
\address[org1]{Department of Electrical Engineering and Computer Science, University of California, Irvine, CA 92697, USA}
\begin{document}

\newmaketitle

\begin{abstract}
Reconfigurable intelligent surfaces (RISs), an emerging technology proposed for inclusion in next generation wireless communication systems, are programmable surfaces that can adaptively reflect incident electromagnetic radiation in different desired directions. To reduce the complexity and physical profile of conventional RIS designs, a novel concept known as Wave-Controlled RIS has been proposed, in which standing waves along a transmission line are used to generate the required dc bias for reflective control. This paper shows the design of such a Wave-Controlled RIS and its biasing transmission line. The effectiveness of this approach in generating the correct dc bias from a single standing wave frequency is analyzed through both theoretical modeling and experimental validation, which uncovered a dependence on impedance matching not accounted for by the theory. Additionally, the potential for reflective control using only a single standing wave frequency on the biasing transmission line is explored, demonstrating the ability of single-beam steering toward angles near broadside.

% We examine the effectiveness of a transmission line design to generate dc voltage for biasing varactors on a reconfigurable intelligent surface (RIS). Our approach involves utilizing multiple biasing standing waves (BSW) along a biasing transmission line (BTL) situated beneath the RF reflective surface....

\end{abstract}

\section{Introduction}
%Reconfigurable intelligent surfaces (RIS) are...

%- Include the definition and applications of RIS, the problem of biasing multiple elements, references to our previous papers...

%Reconfigurable intelligent surfaces (RIS) are composed of numerous unit cells, each incorporating electrically tunable components like varactors, which act as variable capacitors \cite{Costa21}. These components can be dynamically adjusted in real time to manipulate incoming electromagnetic waves, enhancing signal strength and optimizing the wireless communication environment. A novel method for biasing varactors in reconfigurable intelligent surfaces (RIS) has been recently proposed, utilizing standing waves on a transmission line (TL) to bias each RIS element, thereby eliminating the need for a complex external biasing network \cite{Ayanoglu22}...

The implementation of reconfigurable intelligent surface (RIS) technology presents a novel way to improve wireless communication performance by providing adaptive degrees of freedom to design radio-frequency (RF) channels. Using an RIS, signals can be steered around obstacles, beamforming gains can be increased to boost signal-to-noise ratio (SNR) and lower interference, and the overall quality of service (QoS) experienced by network users can be greatly enhanced \cite{Sharma_Reconfigurable_21}. RIS technology is especially beneficial in challenging RF environments like urban areas and indoor spaces, where it can be difficult to sustain strong and reliable signal transmissions~\cite{Abeywickrama20}.

An RIS consists of a wide range of reconfigurable subwavelength components, an example of which would be metallic patches, that are electronically controlled to manipulate electromagnetic waves \cite{Costa21}. These unit cells have the ability to precisely control how incoming electromagnetic waves behave by dynamically modifying their reflection characteristics in real time using tunable elements such as PIN diodes, varactor diodes, and liquid crystals, among others \cite{Yang_Beyond_24}. RISs can increase signal strength, boost wireless network performance, and open the door for more effective communication systems by optimizing wave reflection and propagation.

RIS designs can have hundreds to thousands of unit cells operating at millimeter-wave and terahertz frequencies, resulting in notable beamforming gains and allowing for highly directed and reconfigurable beams \cite{Du_Millimeter_21, cui_coding_14}. Although these qualities improve communication, there are several obstacles to overcome when implementing large-scale RIS designs. Most RIS systems depend on external control units, such as field-programmable gate arrays (FPGAs) or microcontrollers, to configure the reconfigurable elements on the RIS surface \cite{Basar_Wireless_19}. For an RIS with $M$ elements, these control units perform optimization processes involving $M$ variables and transmit the optimal configuration to each of the $M$ RIS elements. Thus, each unit cell requires a dedicated connection to receive the appropriate control voltages or currents, leading to complex wiring networks with hundreds of individual signal connections. This not only complicates fabrication but also increases hardware costs and limits practical deployment. Alternative control strategies such as light-based approaches employing photodiodes, scanning lasers, or encoded light images, have been investigated in order to overcome these difficulties \cite{Hershko_Photodiode_24,Sayanskiy_2D-Programmable_23,Miao_Light-Controlled_23,Hu_Ultrafast_19,Zhang_Light-Controllable_18}. Despite their potential, these methods face limitations, including susceptibility to environmental factors such as rain and reduced resolution due to the distance between the light source and the RIS.

%The majority of RIS systems rely on external control blocks, like FPGAs or other microcontrollers, to configure themselves because they lack active elements \cite{Yuan21}. The best configuration, which includes $M$ values, must be transmitted to the RIS elements by these control blocks, which carry out complex optimization processes involving $M$ variables.

%RIS systems' scalability presents further difficulties because each unit cell normally needs a separate connection in order to receive control voltages or currents. This limits feasible deployment, complicates fabrication, and raises hardware prices by creating complex wiring networks with hundreds of individual signal connections \cite{SBYLSGL23}. Alternative control strategies, like light-based approaches employing photodiodes, scanning lasers, or encoded light images, have been investigated in order to overcome these difficulties \cite{Hershko_24,Sayanskiy_23,Miao_23,Hu_19,Zhang_18}. These methods do have drawbacks, though, such as being vulnerable to weather-related issues like rain and having resolution limitations because of the distance between the light source and the RIS.

\begin{figure}[htpb]
\centering
\includegraphics[width=1.01\columnwidth]{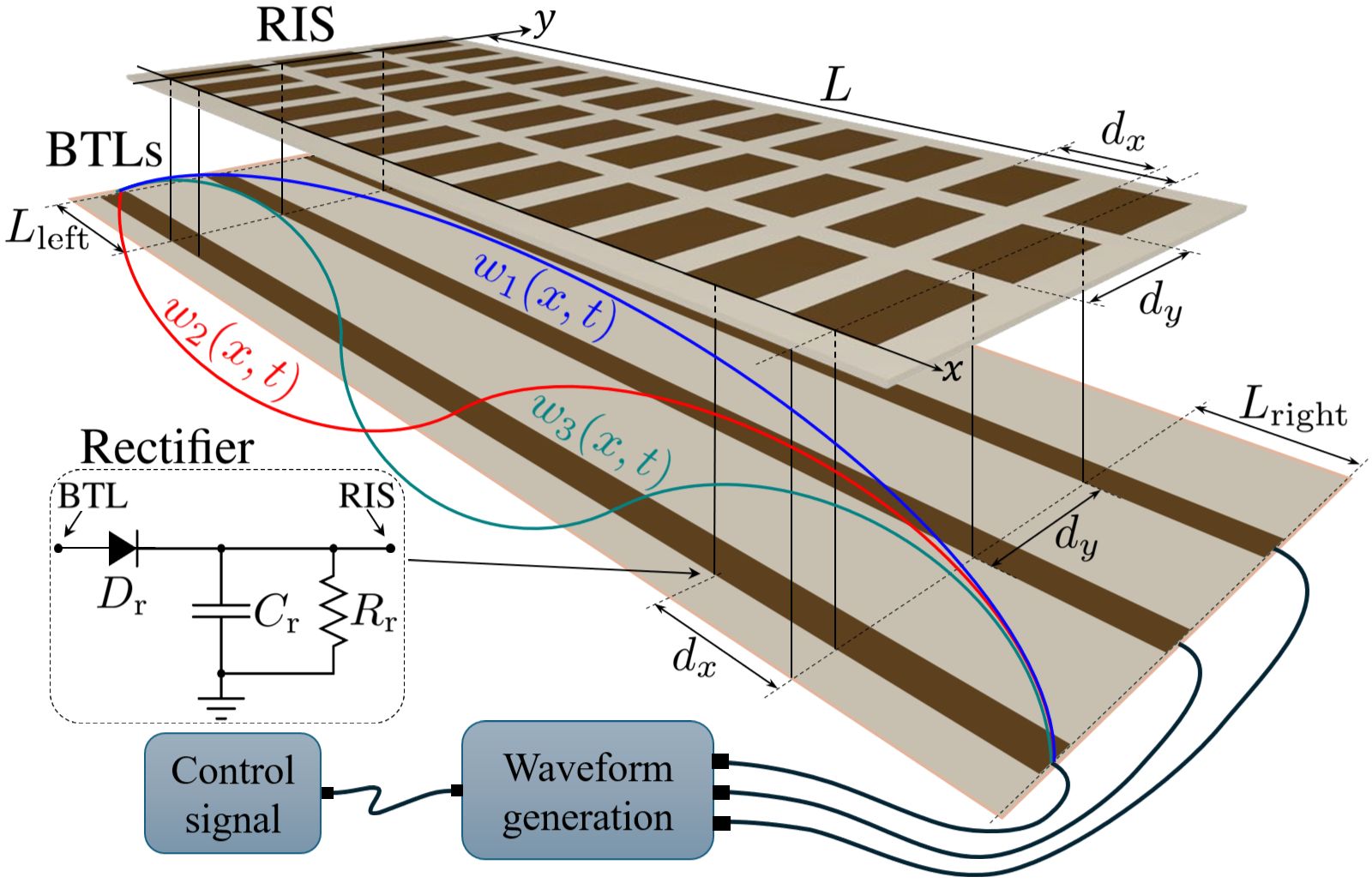}
%\vspace{-0.5cm}
\caption{ Wave-Controlled RIS System: The top layer consists of $M$ RIS elements, which are metallic patches connected to varactors uniformly spaced by $d_x$ along the $x$ direction in each row. The bottom layer contains $N$ biasing standing waves $w_n(x,t)$ that are excited on each BTL. Rectifiers uniformly spaced by $d_x$ in the $x$ direction are used to generate the dc biasing voltages for the varactors. Each BTL is controlled through only one connection on the right, where the $N$ frequencies are injected. Here, a BTL is straight just for conceptualization, though in practice we use a meandered line to be able to use very low-frequency BSWs.}
\label{fig:Wave_Controlled_RIS_General}
%\vspace{-0.4cm}
\end{figure}

This paper investigates the performance of the technique presented in \cite{Ayanoglu22} and later studied in \cite{Saavedra-Melo_APS_RIS_23, Saavedra-Melo_APS_RIS_24, Ben-Itzhak24, Saavedra-Melo_LACAP_RIS_24, Ben-Itzhak2025AI-Driven} as a realistic and effective way to control RIS elements. A schematic is shown in 
Fig.~\ref{fig:Wave_Controlled_RIS_General}. This technique, utilizing varactors for the reconfigurable reflective control, consists of delivering dc biasing voltages to the varactors inside the unit cells that are connected at specific positions, $x_m$, along a single biasing transmission line (BTL). The desired dc biasing voltages supplied to every RIS element to achieve a specific reflective radiation pattern, $w(m)$, can be obtained from combinations of $N$ biasing standing waves (BSWs), $w(x_m, t)$, that are injected into the BTL. The scope of this paper is limited to the study of the performance of the BTL when only a single frequency is injected into the BTL. Further, the extent to which a single reflected beam can be controlled through this single biasing frequency is explored. This technique drastically lowers signaling overhead and hardware complexity by doing away with the requirement for separate wired connections to every unit cell. Advanced RIS functions, including beam creation, polarization control, and channel equalization, are made possible by this straightforward yet efficient control method, which makes it a viable option for large-scale RIS deployments.

\section{Biasing Standing Waves} 
To use a wave-controlled feed network for the RIS, an analytical model for the voltages was derived from a transmission line analysis of the system. This model determines the control voltages that are applied to the unit cells by the input wave, and thus is necessary for understanding how the unit cells are reconfigured. 

%This reconfigurability allows for the optimization of the communication channels between users.
%\subsection{Theoretical Description of the Model}
%- Include the definition of additional parameters and variables (other than the ones mentioned in the Sec. of TL design)...

The wave-controlled RIS approach involves two key components: a BTL board and an RIS board, as shown in 
Fig.~\ref{fig:Wave_Controlled_RIS_General}. The BTL board, placed beneath the reflective surface and shielded from RF interference by a ground plane, supports $N$ BSWs. Rectifiers placed at $M$ positions on the BTL board generate dc voltages, $w(m)$, from these waves which are used to bias the $M$ varactors positioned on the RIS board at $x_m$ for $m=0,1,\ldots,M-1$. This process modifies the reflected phase distribution, enabling control of the RIS’s scattered radiation pattern \cite{Ben-Itzhak24}. The necessary voltage applied to each varactor, $w(m)$, for a given reflected phase is calculated using an analytical model that accounts for the RIS’s reflection behavior under various dc bias conditions for the varactors \cite{Hanna22, Ben-Itzhak24}. Once the required voltages are determined, they are generated by a periodic time-domain signal comprising a combination of $N$ BSWs with specified amplitudes $W_n$, for $n=1,2,\ldots, N$, that are injected into the BTL through a single input port, along with a base voltage $W_0$. The inclusion of $W_0$ simplifies the reconstruction of the voltage distribution, as the varactor's biasing voltage range typically does not include 0 V. Consequently, the expected voltage distribution pattern includes a dc component.

As seen on the RIS board in Fig.~\ref{fig:Wave_Controlled_RIS_General}, the RIS unit cells are separated by a distance $d_x$ along the $x$ direction. Likewise, the corresponding rectifiers that convert the standing wave amplitudes, $w_n(x_m,t)$, into dc values for each RIS unit cell are also placed a distance $d_x$ away from each other along the $x$ direction on the BTL board. These, along with the voltage offset $W_0$, produce the necessary voltages $w(m)$ \cite{Ben-Itzhak24}. The position along the BTL of the $m$th rectifier element is consequently defined as $x_m = md_x$.

The BTL board is not necessarily as simple as that shown in Fig.~\ref{fig:Wave_Controlled_RIS_General}, an example of which is provided in \cite{Saavedra-Melo_APS_RIS_24}. This BTL must connect with the RIS board at specific positions along the $x$ direction within the length $L$, where $L = (M-1)d_x$. The total length of the BTL along the $x$ direction is thus defined as $L_{\mathrm{tot}} = L+L_{\mathrm{left}}+L_{\mathrm{right}}$, where $L_{\mathrm{left}}$ and $L_{\mathrm{right}}$ are additional segments on either end of the BTL 
% An example of the TL board is provided in \cite{Saavedra-Melo_APS_RIS_24}. The BTL extends along the $x$-axis with a total length of $L_{\mathrm{tot}} = L+L_{\mathrm{left}}+L_{\mathrm{right}}$, where $L = (M-1)d_x$, $d_x$ represents the spacing between adjacent RIS elements, and $L_{\mathrm{left}}$ and $L_{\mathrm{right}}$ are additional segments on either end of the BTL. 
that help refine the voltage distribution pattern along the $M$ elements.
The phase velocity of the propagated modes projected along the $x$ direction is $v_\mathrm{b} = c/n_{\mathrm{slow}}$, where $c$ represents the speed of light and $n_{\mathrm{slow}} \ge 1$ is a slowness factor determined by the material properties and geometry of the BTL. The injected periodic time-domain signal oscillates at the biasing frequency $f_\mathrm{b} = v_\mathrm{b}/\lambda_\mathrm{b}$, where $\lambda_\mathrm{b}=2\pi /k_\mathrm{b}$ is the guided wavelength, and the corresponding propagation wavenumber is $k_\mathrm{b} = 2\pi f_\mathrm{b}/ v_\mathrm{b}$.

%Assuming short circuit termination at one end of the BTL, considering $N$ biasing modes (degrees of freedom to control the RIS response) and a phase $\phi_n$ of the $n$th biasing mode, the biasing voltage of the standing wave before rectifying it is

%\vspace{-0.5cm}
%\begin{equation}
%w(x_m,t) = W_0 + \sum_{n=1}^{N} W_n  \sin\left(n_{\mathrm{gs}} k_{b,n} x_m\right) \sin\left(2 \pi f_{b,n}  t + \phi _n \right).
%\label{eq:biasTLvoltage}
%\end{equation}
%\vspace{-0.6cm}

% Here's a Draft I was working on (Ben) for this section, still pretty Rough

%It is important from this design to be able to model the voltages of the BSW at or after each rectifier for this technology to be useful. One can derive the voltage of the standing wave at any point along the BTL simply by applying transmission line theory \cite{pozar2012}. Assuming the lossless case and a single standing wave excited along the $x$-axis, the general solution to this problem is $\textbf{V}(x) = Ae^{-j\beta x}+Be^{j\beta x}$, where $\beta$ is the propagation constant, $A$ is the magnitude of the phasor of the incident wave and $B$ is the magnitude of the phasor of the reflected wave. The short-circuited boundary conditions give a reflection coefficient of $\Gamma_0=-1$, which, when applied to the solution gives $B = -A$. Allowing for a complex amplitude the time-dependent solution for this simplifies to $V(x,t)=2A \mathrm{sin}(\beta x)\mathrm{sin}(\omega_0 t + \theta)$ for a single frequency. 

This design approach enables modeling of the voltage of the BSW before and after each rectifier. The voltage of the standing wave at any point along the BTL can be derived by applying transmission line theory \cite{pozar_microwave_2012_ch2}. In the general case when there are $N$ biasing modes (degrees of freedom to control the RIS response), the biasing voltage of the standing wave before rectification and assuming a short circuit termination at $x=-L_{\mathrm{left}}$ is \cite{Ayanoglu22, Ben-Itzhak24, Ben-Itzhak2025AI-Driven}
\begin{comment}
%\vspace{-0.5cm}
\begin{equation}
\begin{split}
    & w(x,t) = \\
    &
W_0 + \sum_{n=1}^{N}W_n \sin \left( \frac{n \pi \left( x+L_{\mathrm{left}}\right)}{L_{\mathrm{tot}}} \right) \sin \left(n \omega_{\mathrm{b}}  t + \phi_n \right).
\end{split}
\label{eq:StdWave_Volt}
\end{equation}
%\vspace{-0.6cm}
%\vspace{-0.5cm}
\begin{equation}
\begin{split}
& w(x, t) = \\
& W_0 + \sum_{n=1}^{N}W_n  \sin\left(
\frac{ n\omega_\mathrm{b} n_{\mathrm{slow}}}{c}
\left(x+L_\mathrm{left}\right)\right) \sin\left(n \omega_{\mathrm{b}}  t + \phi_n \right).
\end{split}
\label{eq:StdWave_Volt}
\end{equation}
%\vspace{-0.6cm}
\end{comment}
%\vspace{-0.5cm}
\begin{equation}
\begin{split}
& w(x, t) = W_0 + \\
&  \sum_{n=1}^{N}W_n \sin\left(
\frac{ n\omega_\mathrm{b} n_{\mathrm{slow}}}{c}
\left(x+L_\mathrm{left}\right)\right) \cos\left(n \omega_{\mathrm{b}} t + \phi_n \right).
\end{split}
\label{eq:StdWave_Volt}
\end{equation}
%\vspace{-0.6cm}
Here, $n$ is the biasing mode index and $W_n$ is the real-valued amplitude of the standing wave of the $n$th biasing mode. The term $W_0$ is the dc component, which sets the minimum biasing voltage applied to the BTL and shifts the varactor's operating range. Furthermore, $\phi_n$ is the phase offset of the $n$th biasing mode. This definition generalizes Eq. (7) of Ref. \cite{Ben-Itzhak24}, as here the wavelength $\lambda_\mathrm{b}$ and, consequently, the fundamental frequency $f_\mathrm{b}$, do not necessarily depend on the total length of the BTL $L_\mathrm{tot}$; in other words, we do not need standing waves that are resonant with the BTL length.

The $m$th rectifier, shown in Fig.~\ref{fig:Wave_Controlled_RIS_General}, acts as a peak detector and is described in detail in Sec. III. Its output voltage is given by 
\begin{comment} 
\begin{equation}
\begin{split}
    & w(x_m) = \\
    &
\max_t \left[ W_0 + \sum_{n=1}^{N}W_n \sin \left( \frac{n \pi \left( x_m+L_{\mathrm{left}}\right)}{L_{\mathrm{tot}}} \right) \sin(n \omega_{\mathrm{b}} t + \phi_n) \right].
\end{split}
\label{eq:BTL_Volt_Retif}
\end{equation}
\begin{equation}
\begin{split}
    & w(x_m) = W_0 + \\
    &
\max_t \left[\sum_{n=1}^{N}W_n \sin\left(
\frac{ n\omega_\mathrm{b} n_{\mathrm{slow}}}{c}
\left(x_m+L_\mathrm{left}\right)\right) \sin(n \omega_{\mathrm{b}} t + \phi_n) \right].
\end{split}
\label{eq:BTL_Volt_Retif}
\end{equation}
\end{comment}
\begin{equation}
\begin{split}
    & w(m) = W_0 + \\
    &
\max_t \left[\sum_{n=1}^{N}W_n \sin\left(
\frac{ n\omega_\mathrm{b} n_{\mathrm{slow}}}{c}
\left(x_m+L_\mathrm{left}\right)\right) \cos(n \omega_{\mathrm{b}} t + \phi_n) \right].
\end{split}
\label{eq:BTL_Volt_Retif}
\end{equation}
%Interestingly, due to the periodicity and symmetry of the problem, the actual length of the line has no effect on the output voltage, only the position along that length. 
In \cite{Ayanoglu22, Ben-Itzhak24}, we have explored theoretically the use of the BTL with several standing waves. Here, we focus on the experimental characterization of a BTL and on the use of a single BSW to control the pointing angle of a single beam. Accordingly, when using a single BSW of frequency  $f_\mathrm{b}$, (\ref{eq:StdWave_Volt}) reduces to
%\vspace{-0.5cm}
\begin{equation}
\begin{split}
    & w(x, t) = W_0 +  \\
    & W_\mathrm{b} \sin\left(
\frac{\omega_\mathrm{b} n_{\mathrm{slow}}}{c}
\left(x+L_\mathrm{left}\right)\right) \cos\left(\omega_\mathrm{b} t + \phi \right),
\label{eq:StdWave_Volt_Single}
\end{split}
\end{equation}
%\vspace{-0.6cm}

\noindent where $W_\mathrm{b}$ is the amplitude and $\phi$ is the phase shift. After rectification at each $x_m$, the dc voltage is 
%\vspace{-0.5cm}
\begin{equation}
w(m) = W_0 + \left|W_\mathrm{b} \sin\left(
\frac{\omega_\mathrm{b} n_{\mathrm{slow}}}{c}
\left(x_m+L_\mathrm{left}\right)\right) \right|.
\label{eq:BTL_Volt_Retif_Single}
\end{equation}
%\vspace{-0.6cm}

%\subsection{Considerations Due to the Standing Waves}
%- Describe how the last version of the model was obtained to model accurately the performance of the standing waves present along the TL (time dependency...)...

\begin{comment}

\subsection{Optimization Process}
For beamforming applications, there is a necessary voltage desired at each rectifier $v_m$. In order to match the desired voltages to those output by the model a simple L2 Norm cost function can be defined to do so,

\begin{equation}
J = \sum(v_m^2-w_m^2).
\label{eq:biasTLvoltage}
\end{equation} 

As these waves are induced by an arbitrary waveform generator, the parameters able to be optimized are the harmonic amplitudes and phase constants. However, as the standing waves of this function are non-convex, the maximizing function is not necessarily convex and has been found to be non-differentiable, though still continuous \ref{Maybe ?? Figure}.

This leads to difficulties in the optimization, and necessarily local search techniques are required to solve. Several unique techniques have been applied to this problem \cite{itzhak2024design}.
\end{comment}

\section{System Design}

The wave-controlled RIS system is realized here using two separate boards: the RIS board and the BTL board. The RIS board reflects the incident radiation based on a spatial distribution of the phase reflection coefficient across the RIS elements, which is governed by the biasing voltage distribution $w(m)$ provided by the BSW, supported by the BTL board. Other implementations could be realized in a single multilayer board. This biasing pattern is controlled by the signal frequencies and amplitudes injected into the BTL (see Fig.~\ref{fig:Wave_Controlled_RIS_General}).
%on the control scheme, i.e. the biasing voltage distribution, provided by the BTL board. 
The RIS unit cell is designed to achieve the desired phase response by applying a variable voltage to the varactor. 
%in a given RF frequency range.
The design includes the unit cell's geometry, material, and chosen varactor. The BTL board comprises a transmission line with a defined geometry and integrated rectifiers to provide the dc biasing voltage for the unit cells of the RIS. Material specifications play a crucial role in the overall performance of the BTL and are introduced next.

\subsection{RIS Unit Cell Design}

\begin{figure}[htpb]
\centering
\includegraphics[width=1.01\columnwidth]{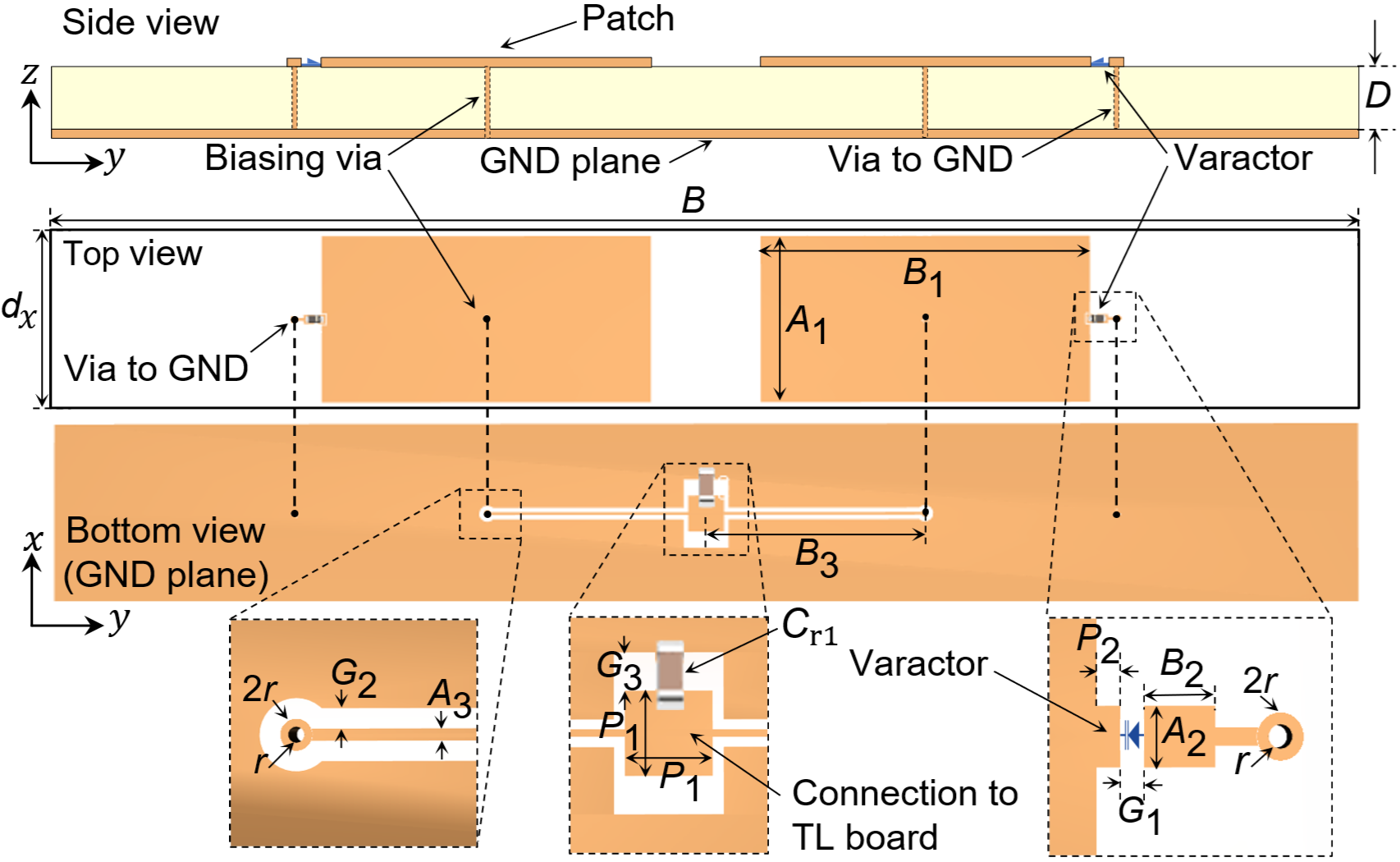}
%\vspace{-0.5cm}
\caption{ RIS board unit cell. Two rectangular patches are arranged in a mirrored configuration on a grounded dielectric substrate. Each patch is connected to the cathode of an individual varactor with the varactor's anode grounded via a through-hole connection. The varactors are biased using a dc voltage provided by a via between the center of each patch and a coplanar waveguide with ground that is itself connected to the BTL board (not shown here).}
\label{fig:RIS_Dimensions.png}
%\vspace{-0.4cm}
\end{figure}

The RIS unit cell consists of an array of two rectangular patches arranged in a mirrored configuration as in Fig.~\ref{fig:RIS_Dimensions.png}, mounted on a grounded dielectric substrate (RT5880LZ) with a relative permittivity of $\epsilon_r = 2$, a loss tangent of $\tan \delta = 0.0021$, and a thickness of 1.27 mm. Each patch is connected to the cathode of an individual SMV1231-040LF varactor, while the varactor's anode is grounded via a through-hole connection. We assume an incident $y$-polarized electric field, and the unit cell design is optimized to maximize the phase dynamic range within the industrial, scientific, and medical (ISM) 2.45 GHz band for reverse bias voltages ranging from $4$ V to approximately $10$ V.  

The rectified reverse voltage $w(m)$ is provided to a varactor on the top RIS layer in the $m$th unit cell through a vertical biasing via from the center of the RIS patch to a coplanar waveguide on the ground plane of the RIS board (Fig.~\ref{fig:RIS_Dimensions.png}). This line is oriented to avoid disrupting the RIS ground-plane current. This configuration allows efficient biasing of the varactors while minimizing the effect on the RIS RF current flow. To further suppress the effect on the RF current at the connection node located at the center of the coplanar waveguide, the capacitor $C_{r1}=100$ pF is incorporated to ensure proper grounding at RF (Fig.~\ref{fig:RIS_Dimensions.png}). This capacitor is also part of the rectifier in the BTL board, as discussed in Sec. III B. The RIS board comprises 27 identical unit cells with two copper patches, with each unit cell featuring the following dimensions (in mm): $d_x = 20$, $A_1 = 18.5$, $A_2 = 1$, $A_3 = 0.3$, $B = 145.4$, $B_1 = 36.7$, $B_2 = 1.2$, $B_3 = 24.8$, $G_1 = 0.4$, $G_2 = 0.5$, $G_3 = 1.8$, $r = 0.2$, $P_1 = 4$, $P_2 = 0.4$, as shown in Fig.~\ref{fig:RIS_Dimensions.png}. 

\begin{figure}[htpb]
\vspace{-0.3cm}
\centering
\includegraphics[width=0.7\columnwidth]{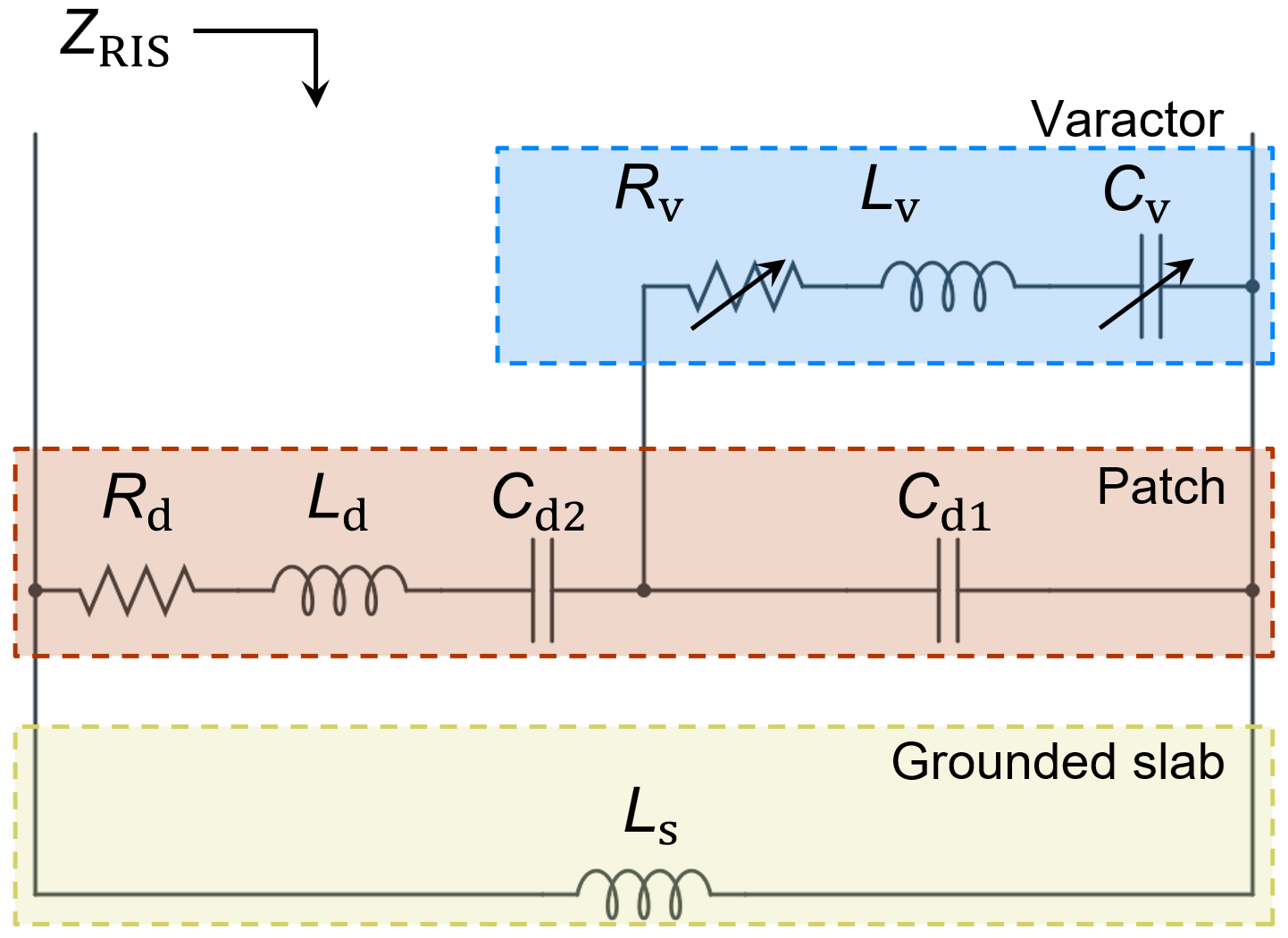}
%\vspace{-0.35cm}
\caption{Analytic circuit model to calculate the RIS impedance, $Z_{\mathrm{RIS}}$, seen from a plane wave impinging from the top. Since this model accounts for the effect of the tunable varactor, the impedance $Z_{\mathrm{RIS}}$ is a function of the biasing voltage $V$.}
\label{fig:Analytical_Model_RIS}
\vspace{-0.2cm}
\end{figure}

%To determine the tunable impedance 
To calculate the reflected phase provided by each RIS element, the analytical circuit model shown in Fig.~\ref{fig:Analytical_Model_RIS} is summarized in \cite{Hanna22, Ben-Itzhak24, Ben-Itzhak2025AI-Driven}. This model requires only one full-wave simulation of the unit cell over the frequency of operation (assuming the varactor is disconnected) to extract the circuit elements that represent just the patch over the grounded slab, as detailed in Appendix A. For the design presented in Fig.~\ref{fig:RIS_Dimensions.png}, the values obtained are $L_\mathrm{s}=1.60$~nH, $R_\mathrm{d}=0.17$ $\Omega$,  $L_\mathrm{d}=1.92$ nH, and $C_\mathrm{d}=0.98$ pF represents the two fringe capacitances at the two sides of each patch toward the ground and toward adjacent patch elements, as shown in Fig. 3 of \cite{ZhangIEEETAP2003PlanarAMC} and Fig.13 of \cite{donzelli2009metamaterial}.% The details of the retrieval of these parameters are provided in [??] and hence are not repeated here. 

%these elements include a resistance $R_{\mathrm{d}}=0.1671$ $\Omega$, a capacitance $C_{\mathrm{d}}=0.97821$ pF, and an inductance $L_{\mathrm{d}}=1.9177$ nH. 

Once these circuit elements are determined, the varactor is incorporated to complete the RIS unit cell model by adding its RF impedance $Z_\mathrm{v}$ to the equivalent circuit as shown in Fig.~\ref{fig:Analytical_Model_RIS}. In this configuration, the varactor is physically connected only on one side of the patch toward the ground, and therefore its impedance $Z_\mathrm{v}$ is in parallel to only one of the two fringe capacitors. Accordingly, the lumped capacitor $C_\mathrm{d}$ is divided into two series capacitors and, after tuning, the fitting with full-wave results provides the values $C_\mathrm{d1}=1.25$ pF and $C_\mathrm{d2}=2.72$ pF. We have verified that the model in Fig.~\ref{fig:Analytical_Model_RIS} with split capacitors $C_\mathrm{d1}$ and $C_\mathrm{d2}$ produces more accurate results than considering a single capacitor $C_\mathrm{d}$ in parallel with the varactor impedance, over the whole range of frequencies and varactor voltages considered. The circuit in Fig.~\ref{fig:Analytical_Model_RIS} replaces Fig. 4 in \cite{Ben-Itzhak2025AI-Driven}, where the splitting of the capacitance $C_{\text{d}}$ into two separate capacitors, $C_{\text{d1}}$ and $C_{\text{d2}}$, was omitted from the figure but accounted for in the numerical results.

The varactor's impedance is modeled by an equivalent series RLC circuit
\begin{equation}
Z_\mathrm{v}(V)=R_{\mathrm{v}}(V) + j\omega L_{\mathrm{v}} + \frac{1}{j\omega C_{\mathrm{v}}(V)},   
\end{equation}
where the variable capacitance $C_{\mathrm{v}}\mathrm{(V)}$ and variable resistance $R_{\mathrm{v}}\mathrm{(V)}$ are functions of the reverse biasing voltage $V$, and the constant inductance $L_{\mathrm{v}}=2.39$ nH models both the package inductance and the parasitic inductance introduced when the varactor is soldered across the unit cell gap. The values of $C_{\mathrm{v}}\mathrm{(V)}$ and $R_{\mathrm{v}}\mathrm{(V)}$ are obtained through a parametric sweep simulation in the commercial Keysight Advanced Design System (ADS) software, based on the SPICE model of the varactor provided by the manufacturer. These values are limited to the reverse voltage bias range of $[4$ V, $15$ V$]$ and are detailed in Table \ref{tab:Cv_Rv_Varactor}.

%\begin{table}[]
%\caption{Values of $C_\mathrm{v}$ and $R_\mathrm{v}$ of the varactor model in Fig. \ref{fig:Analytical_Model_RIS} for different values of the varactor biasing voltage.}
%\centering
%\begin{tabular}{|l|l|l|}
%\hline
%\multicolumn{1}{|c|}{$\mathrm{V}$ (V)} & \multicolumn{1}{c|}{$C_\mathrm{v}$ (pF)} & \multicolumn{1}{c|}{$R_\mathrm{v}$ ($\Omega$)} \\ \hline
%-15 & 0.460 & 0.005 \\ \hline
%-14 & 0.465 & 0.007 \\ \hline
%-13 & 0.471 & 0.011 \\ \hline
%-12 & 0.478 & 0.016 \\ \hline
%-11 & 0.488 & 0.024 \\ \hline
%-10 & 0.501 & 0.037 \\ \hline
%-9  & 0.519 & 0.058 \\ \hline
%-8  & 0.544 & 0.091 \\ \hline
%-7  & 0.578 & 0.142 \\ \hline
%-6  & 0.626 & 0.221 \\ \hline
%-5  & 0.697 & 0.340 \\ \hline
%-4  & 0.802 & 0.509 \\ \hline
%\end{tabular}
%\label{tab:Cv_Rv_Varactor}
%\end{table}

\begin{table}[]
\caption{Values of $C_\mathrm{v}$ and $R_\mathrm{v}$ for the varactor model in Fig.~\ref{fig:Analytical_Model_RIS} for different values of the varactor biasing voltage.}
\centering
\begin{tabular}{|l|l|l|}
\hline
\multicolumn{1}{|c|}{$\mathrm{V}$ (V)} & \multicolumn{1}{c|}{$C_\mathrm{v}$ (pF)} & \multicolumn{1}{c|}{$R_\mathrm{v}$ ($\Omega$)} \\ \hline
4  & 0.802 & 0.509 \\ \hline
5  & 0.697 & 0.340 \\ \hline
6  & 0.626 & 0.221 \\ \hline
7  & 0.578 & 0.142 \\ \hline
8  & 0.544 & 0.091 \\ \hline
9  & 0.519 & 0.058 \\ \hline
10 & 0.501 & 0.037 \\ \hline
11 & 0.488 & 0.024 \\ \hline
12 & 0.478 & 0.016 \\ \hline
13 & 0.471 & 0.011 \\ \hline
14 & 0.465 & 0.007 \\ \hline
15 & 0.460 & 0.005 \\ \hline
\end{tabular}
\label{tab:Cv_Rv_Varactor}
\end{table}

As shown in Fig.~\ref{fig:Analytical_Model_RIS}, the RLC varactor model is connected in parallel to the capacitance $C_{\mathrm{d}}$, and the total equivalent impedance of the RIS is hence given by 

\begin{equation}
%\begin{split}
%    & Z_{\text{RIS}}= \\
%    & \left(R_{\mathrm{d}}+j\omega L_{\mathrm{d}} + Z_\mathrm{v}\parallel\frac{1}{j\omega C_{\mathrm{d}}}\right)\parallel j\omega L_{\mathrm{s}}.
%\end{split}
Z_{\text{RIS}}=\left(R_{\mathrm{d}}+j\omega L_{\mathrm{d}} + \frac{1}{j\omega C_{\mathrm{d2}}}+ Z_\mathrm{v}\parallel\frac{1}{j\omega C_{\mathrm{d1}}}\right)\parallel j\omega L_{\mathrm{s}}.
\end{equation}

\begin{comment}
\begin{equation}\label{eq:zin_wm_we}
Z_{\mathrm{RIS}}= \frac{j\omega L_\mathrm{s}\left (1+j\omega R_\mathrm{d}C'_\mathrm{d}-\left (\frac{\omega}{\omega_\mathrm{e}}\right )^{2}\right )}{\left (1+j\omega R_\mathrm{d}C'_\mathrm{d}-\left (\frac{\omega}{\omega_\mathrm{m}}\right )^{2}\right )},
\end{equation}

\noindent where $\omega_\mathrm{e}^2 = 1/\left (C_\mathrm{d}\left (L_\mathrm{d}+L_\mathrm{s}\right )\right )$ and $\omega_\mathrm{m}^2 = 1/\left (C_d L_d\right )$, as described in Appendix A. The parameter $C'_\mathrm{d}$ accounts for the effect of the varactor and is defined as $C'_\mathrm{d}=j\omega Z'_\mathrm{C}$, with $Z'_\mathrm{C}=\left(R_{\mathrm{v}} + j\omega L_{\mathrm{v}} + 1/j\omega C_{\mathrm{v}}\right)\parallel1/\left(j\omega C_{\mathrm{d}}\right)$.

%$Z'_\mathrm{C}=\left(R_{\mathrm{v}} + j\omega L_{\mathrm{v}} + \frac{1}{j\omega C_{\mathrm{v}}}\right)\parallel\frac{1}{j\omega C_{\mathrm{d}}}$
\end{comment}

\begin{figure}[htpb]
%\vspace{-0.3cm}
\centering
\begin{subfigure}{0.5\textwidth}
\includegraphics[width=0.95\columnwidth]{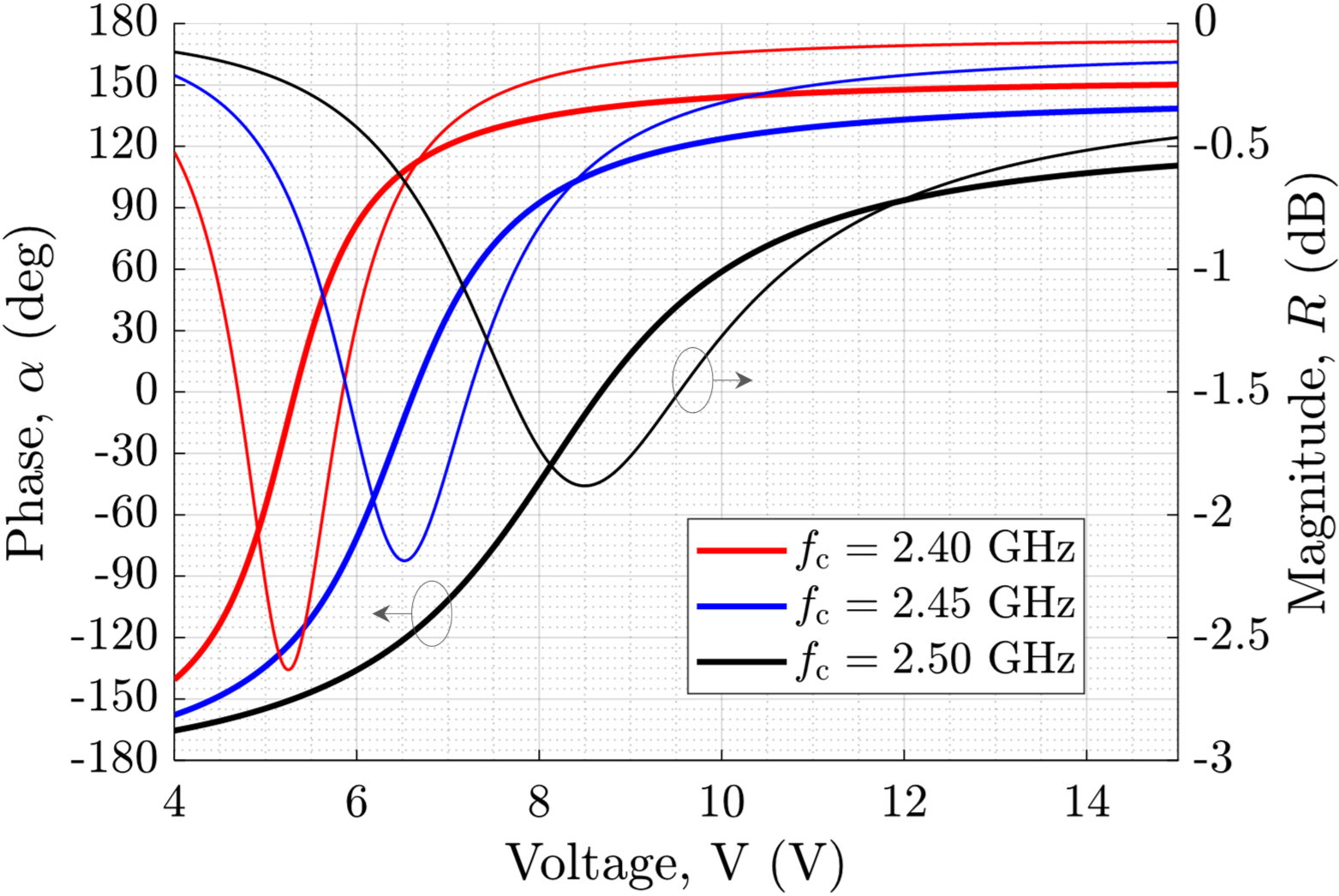}
%\vspace{-0.35cm}
\caption{}
\label{fig:Phase_Mag_vs_voltage}
%\vspace{-0.2cm}
\end{subfigure}
\centering
\begin{subfigure}{0.5\textwidth}
\includegraphics[width=0.95\columnwidth]{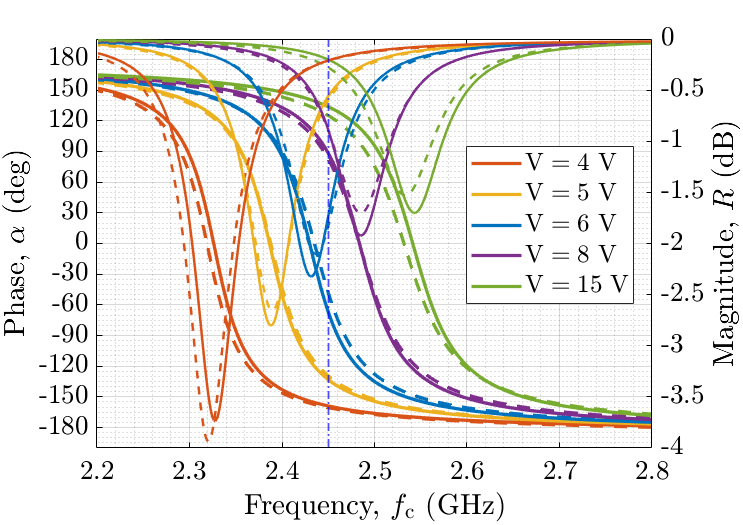}
%\vspace{-0.35cm}
\caption{}
\label{fig:Phase_Mag_vs_frequency}
%\vspace{-0.2cm}
\end{subfigure}
\caption{Phase and magnitude response of the reflection coefficient for RIS unit cells as a function of (a) the varactor's biasing voltage at three different frequencies, based on the analytical model described by (\ref{eq:ReflCoef}), and (b) the frequency at five different biasing voltages. Solid lines represent results from (\ref{eq:ReflCoef}) that are in good agreement with full-wave simulations (dashed lines).}
\label{fig:Phase_Mag_vs_V_freq}
\end{figure}

Finally, $Z_{\text{RIS}}$ is used to determine the reflection coefficients, $\Gamma$, of the RIS elements:
\begin{equation}
    \Gamma = \frac{Z_{\mathrm{RIS}}-\eta_0}{Z_{\mathrm{RIS}}+\eta_0},
\label{eq:ReflCoef} 
\end{equation}
\noindent where $\eta_0=377$ $\Omega$ is the free-space impedance. The magnitude and phase of $\Gamma$ are denoted by $R(\mathrm{V})$ and $\alpha(\mathrm{V})$, respectively. The values of $R$ and $\alpha$ as functions of the biasing voltage for the unit cell in Fig.~\ref{fig:RIS_Dimensions.png} are shown in Fig.~\ref{fig:Phase_Mag_vs_V_freq}(\subref*{fig:Phase_Mag_vs_voltage}) for three different frequencies, including the target RF frequency $f_\mathrm{c}=2.45$ GHz. The voltage $\mathrm{V}$ is positive as the varactor is connected in reverse, with its anode grounded and its cathode connected to the positive biasing voltage. At $f_\mathrm{c}=2.45$ GHz, the phase $\alpha$ varies between $-160^\circ$ and $+140^\circ$, with the most significant variations occurring between $4$ and $10$ V. The lowest reflection magnitude $R$, approximately $-2.4$ dB, is observed around $7$ V. Fig.~\ref{fig:Phase_Mag_vs_V_freq}(\subref*{fig:Phase_Mag_vs_frequency}) illustrates the behavior of $\alpha$ and $R$ as functions of the RF frequency and compares the analytical model (solid lines) with full-wave simulations performed using the commercial software CST Studio Suite (dashed lines) for five different biasing voltages. The agreement is generally good across different biasing voltages and frequencies, with particularly strong accuracy at the target frequency $f_\mathrm{c}=2.45$ GHz (dotted line in blue). Although in practical RIS implementations the control is optimized based on channel estimates, having an analytical model is useful for design and research purposes or to assist the optimization algorithms. 

The model is obtained using periodic boundary conditions for normal incidence. As conventionally done in reflectarrays or RIS metasurfaces, we use the well accepted “local periodicty” approximation: if the $m$th varactor has a voltage bias $w(m)$ at $x_m=md_x$ for $m=0,1,...,M-1$, that is slowly varying with $m$, then the $m$th reflection coefficient $\Gamma_m=R_m e^{j\alpha_m}$ is well approximated by (\ref{eq:ReflCoef}) .

%[DELETE THE RIGHT SIDE IN FIG. 1, ROTATE AND MERGE IT WITH FIG. 2]

\subsection{Biasing Transmission Line Design}
%- Include the definition of all the parameters and variables needed...
%- Include CST simulation results showing performance, s-par losses, geometry, substrate, etc***

The goal is to control the RIS reflection coefficients independently using the BSWs $w_n(x,t)$ in the BTL. After rectification at $x_m=md_x$, control of the dc biasing voltages $w(m)$ leads to control of the reflection magnitudes and phases $R_m$ and $\alpha_m$ on each cell.    

\begin{figure}[htpb]
%\vspace{-0.3cm}
\centering
\includegraphics[width=0.95\columnwidth]{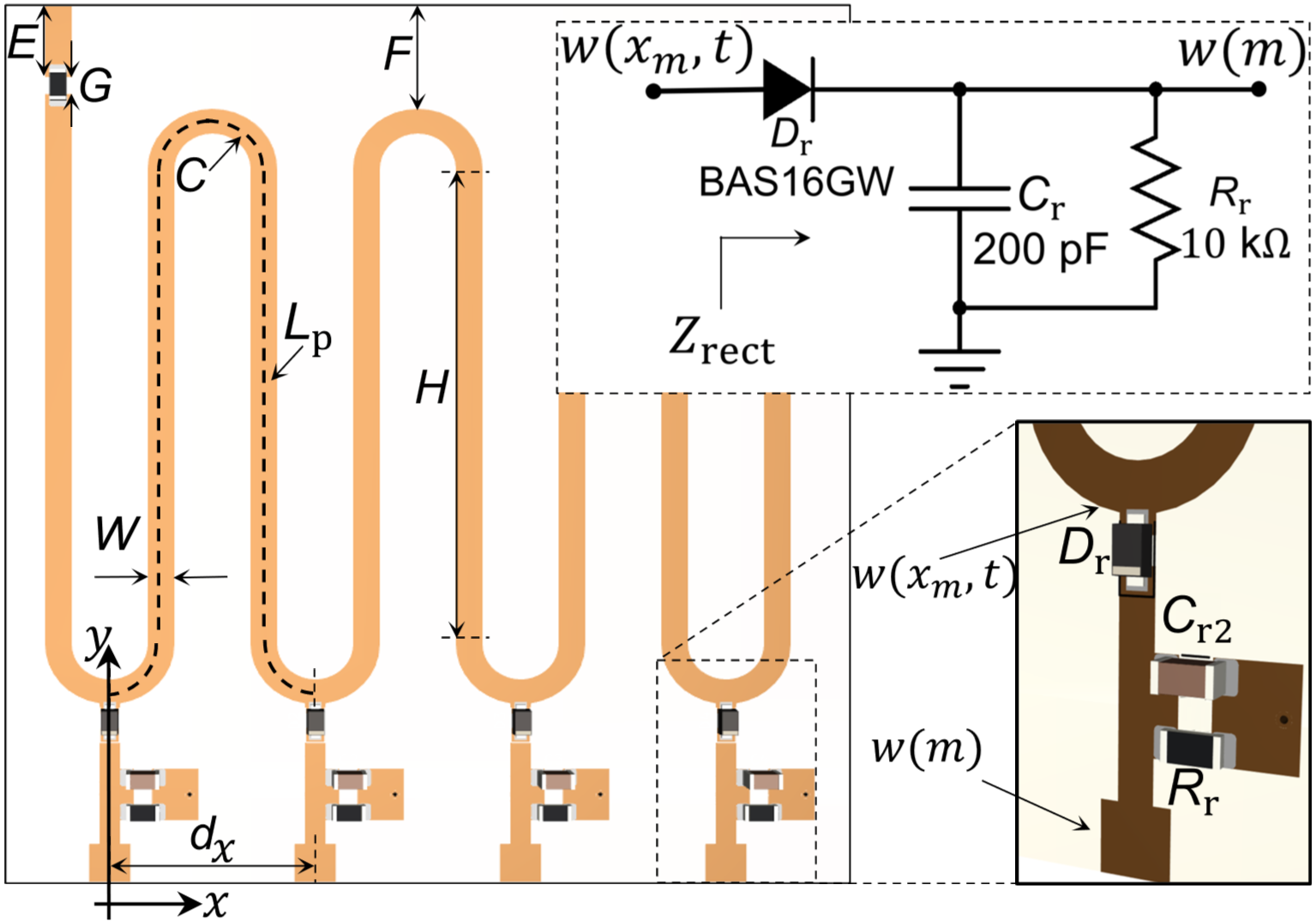}
%\vspace{-0.35cm}
\caption{Geometry of the BTL formed by a meandered microstrip line over a grounded dielectric layer (not shown). The rectifiers of the ac BSW are at the bottom of the figure, one per RIS element. The voltages $w(x_m,t)$, $m=0,1,...,M-1$ are obtained from the lower portion of the BTL. The node of the rectified voltage $w(m)$ is subsequently connected to the coplanar waveguide on the RIS board (connection not shown here).}
\label{fig:TL_geometry_Rectifier}
%\vspace{-0.2cm}
\end{figure}

The BTL is a printed microstrip meandered line (or serpentine), chosen for its compact size, phase delay control, and ease of fabrication; see Fig.~\ref{fig:TL_geometry_Rectifier}. It is made of copper and implemented on a grounded dielectric substrate RO3010 with relative permittivity $\epsilon_r = 11.2$, $\tan \delta = 0.0022$, and thickness $h=0.64$ mm. The BTL board consists of $27$ identical unit cells, where the two cells at the ends have been modified to include SMA connectors that allow connection of the signal generator and integration of possible external loads, if required, to ensure optimal performance within the desired frequency range. The proposed design details and geometry are displayed in Fig.~\ref{fig:TL_geometry_Rectifier}. The dimensions of the BTL unit cell are (in mm): $d_x = 20$, $C = 5$, $E = 7.64$, $F = 11$, $G=1.8$, $H = 50$, $L_{\mathrm{p}} = 131.42$, $W = 2.6$.

Each unit cell of the BTL has a curvilinear path length $L_{\mathrm{p}}$ and a spatial period $d_x$ along the $x$ direction (the same as for the RIS board), and is connected to a rectifier that extracts the desired dc voltage to bias the varactor located on the RIS board. The BTL signal travels along the microstrip %making the meander BTL 
with phase velocity $c/n_{{\mathrm{eff}}}$ where $n_{{\mathrm{eff}}} = \sqrt\epsilon_{{\mathrm{eff}}} = 2.94$ is the effective refractive index of the microstrip and $\epsilon_{{\mathrm{eff}}} = 8.66$ is its effective permittivity, which is approximately calculated as \cite{pozar_microwave_2012_ch3}
\begin{equation}
\epsilon_{\text{eff}} = \frac{\epsilon_r + 1}{2} + \frac{\epsilon_r - 1}{2} \frac{1}{ \sqrt{1 + 12 h/W}}, 
\end{equation}
and is considered dispersionless in the frequency range of operation. The characteristic impedance of the microstrip is $Z_0 = 19.23$ $\Omega$. In this case, the slowness factor, used in (\ref{eq:StdWave_Volt})-(\ref{eq:BTL_Volt_Retif_Single}), is also determined by the path length $L_{\mathrm{p}}$ and the period $d_x$, and is given by $n_{{\mathrm{slow}}} = n_{{\mathrm{geom}}}n_{{\mathrm{eff}}}$, where $n_{{\mathrm{geom}}} = L_{\mathrm{p}}/d_x = 6.57$ represents the geometric index that allows for projection of the standing wave pattern from the printed meandered line onto the $x$ direction. Thus, the resulting slowness factor is $n_{{\mathrm{slow}}} = 19.34$. Although the segments $L_\mathrm{left}$ and $L_\mathrm{right}$ of the BTL follow the same meandered line geometry as the main section $L$, these two additional segments can have a different geometry as they are not connected to any rectifier. However, the values of $L_\mathrm{left}$ and $L_\mathrm{right}$ must be normalized with respect to the geometric index in order to be used in (\ref{eq:StdWave_Volt})-(\ref{eq:BTL_Volt_Retif_Single}). Therefore, the total length of the BTL along the $x$ direction is $L_{\mathrm{tot}} = 540$ mm, with $L = 520$ mm, and $L_\mathrm{left} = L_\mathrm{right} = 0.5d_x = 10$ mm, while the total path length of the meandered microstrip is $3548.3$ mm. 

%$\varepsilon_{\text{eff}} = \frac{\varepsilon_r + 1}{2} + \frac{\varepsilon_r - 1}{2} \cdot \frac{1}{\sqrt{1 + 12 \frac{h}{W}}}$

\begin{figure}[htpb]
%\vspace{-0.3cm}
\centering
\includegraphics[width=0.95\columnwidth]{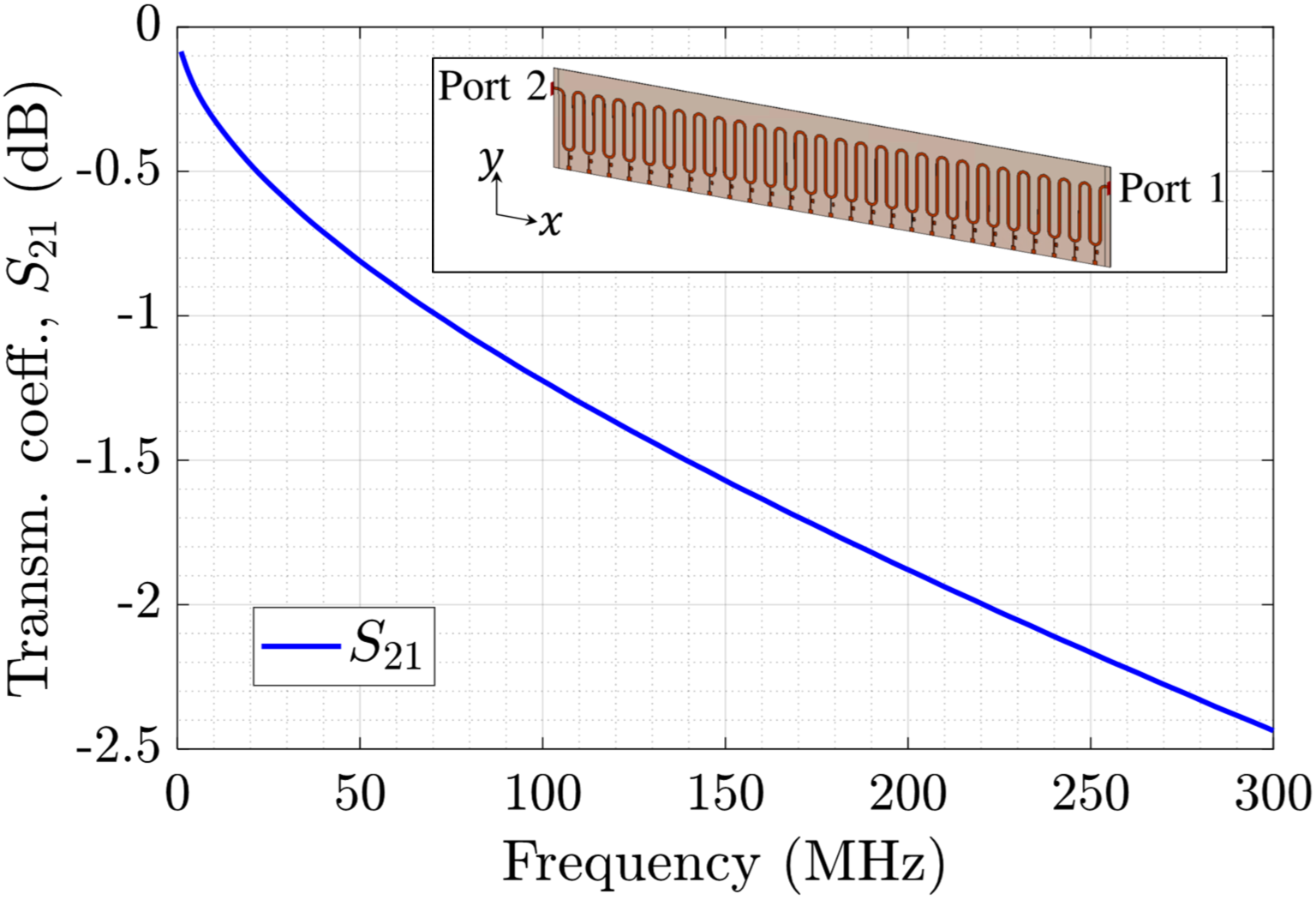}
%\vspace{-0.35cm}
\caption{Transmission coefficient $S_{21}$ of the BTL with $M=27$ elements for a frequency range of 0 to 300 MHz. The value of $S_{21}$ decreases with frequency, reaching $-2.44$ dB at 300 MHz. The lowest frequency of the BSW in this paper is $f_\mathrm{b} = 7.2$ MHz.}
\label{fig:TL_S12_300MHz}
%\vspace{-0.2cm}
\end{figure}

Using printed meandered microstrip topology for the BTL design offers the significant advantage of increasing the slowness of the waves in the $x$ direction, allowing the use of low frequencies for the BSWs. This advantageous feature helps mitigate unwanted interference from the BTL harmonics on the RF system. The scattering parameters have been analyzed to assess the losses and reflection performance of the BTL design. The selected frequency range spans from 0 to 300 MHz, which is much lower than the RF band (here, around 2.45 GHz). When the RIS is used with several standing waves, as in \cite{Ayanoglu22,Ben-Itzhak24}, this design allows for a fundamental frequency of approximately 7.2 MHz. This ensures that at least 21 modes are available below 300 MHz, which is a sufficient number of degrees of freedom for RIS control. The frequency $f_\mathrm{b}$ of the fundamental BSW decreases when the number $M$ of RIS elements increases. Fig.~\ref{fig:TL_S12_300MHz} presents the transmission coefficient of the BTL design (without considering the losses of the 27 rectifiers), obtained from a full-wave simulation in CST Studio Suite. The results show a gradual decrease with frequency due to ohmic and dielectric losses, reaching a minimum value of $-2.44$ dB at 300 MHz. Although not depicted in the figure, the reflection coefficient with Port 2 matched to its characteristic impedance remains consistently below $-34$ dB across the entire frequency range, indicating minimal mismatch. The BSWs are formed along the BTL due to an open or short termination at $x = -L_{\mathrm{left}}$.

%Therefore, the frequency of the fundamental harmonic is $f_{b,1} = 7.19$ MHz.

%\subsection{Rectifiers}

%[MERGE WITH THE PREVIOUS SUBSECTION???]
%- Include theoretical design, time constant, R and C, features of the diode, ADS simulation results in T-D and also F-D

To extract the dc voltages $w(m)$ from the BTL standing wave $w(x_m,t)$ at each $x_m=m d_x$ location, a rectifier circuit composed of a resistor $R_\mathrm{r}$, a capacitor $C_\mathrm{r}$, and a diode $D_\mathrm{r}$ was designed. The capacitor and the resistor are grounded via a through-hole as shown in Fig.~\ref{fig:TL_geometry_Rectifier} (only the microstrip layer is shown, the ground plane underneath is not shown), while the diode is connected between these components and the BTL at position $x_m$. The ADS time-domain circuit simulator was employed to optimize the design of the circuit, resulting in $R_\mathrm{r} = 10$ k$\Omega$ and $C_\mathrm{r} = 200$ pF. To mitigate RF current effects on both boards, the capacitor $C_\mathrm{r}$ is implemented as two 100-pF capacitors in parallel: $C_\mathrm{r1}$ on the RIS board (see Fig. \ref{fig:RIS_Dimensions.png}) and $C_\mathrm{r2}$ on the BTL board. The BAS16GW diode provided by Nexperia was selected for its fast recovery time of 4 ns and low capacitance of 1.5 pF. By choosing a time constant of $\tau=2$ $\mu$s, the rectifier effectively tracks signal peaks while maintaining the impedance $Z_\mathrm{rect}$ seen by the BTL between a few hundred to 1000 $\Omega$ for frequencies up to 300 MHz. The impedance $Z_\mathrm{rect}$ plays a crucial role in the biasing process, as it is connected to the BTL and can influence the voltage distribution along the RIS. However, in this case, the impedance does not significantly load the BTL, and thus does not have a large impact on system performance. Fig.~\ref{fig:TL_geometry_Rectifier} illustrates the rectifier circuit (inset) and its connection to the BTL and the RIS board. 
% (with two parallel 100 pF capacitors placed one on each board)

\begin{figure}[htpb]
%\vspace{-0.3cm}
\centering
\includegraphics[width=1\columnwidth]{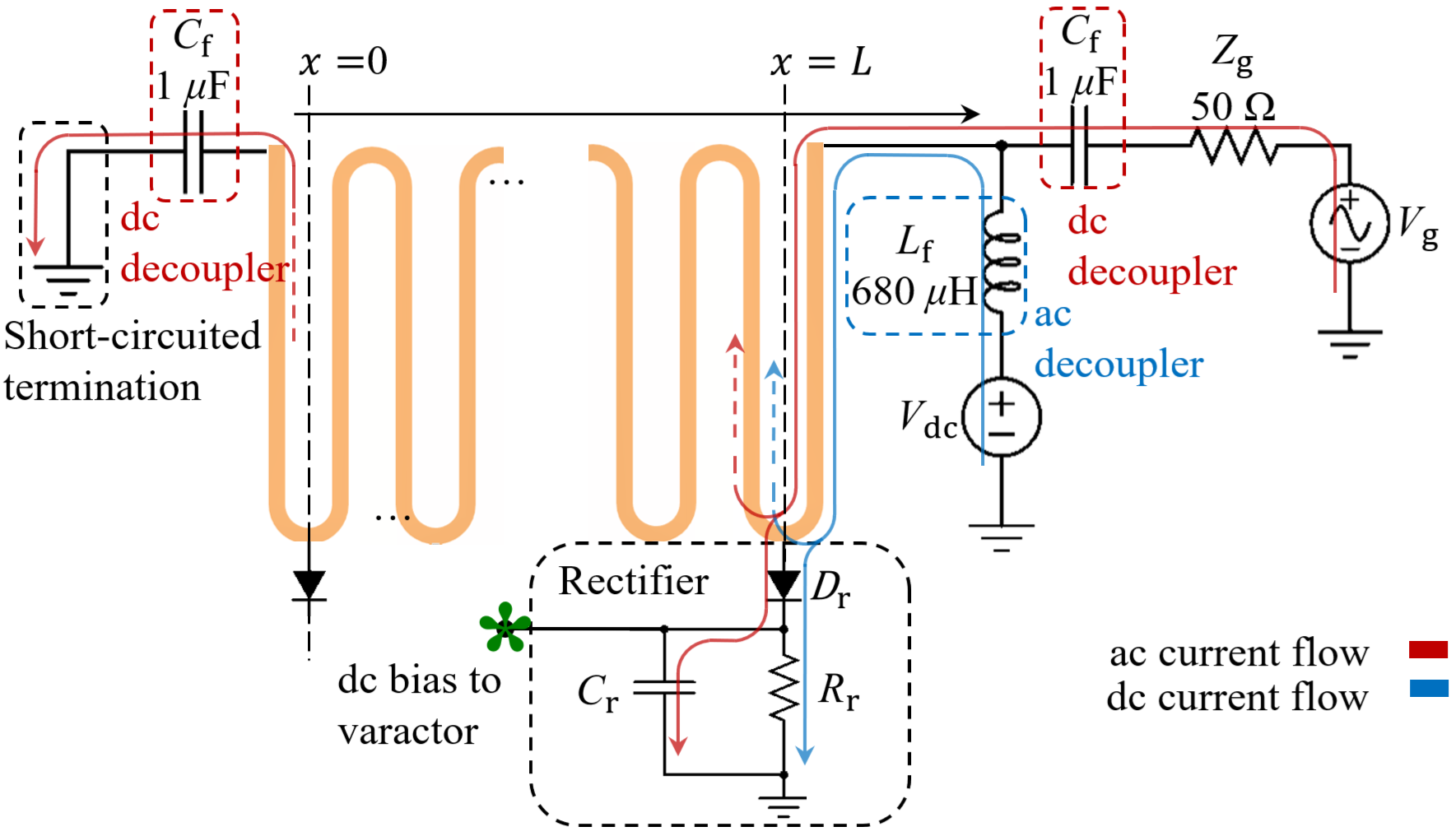}
%\vspace{-0.35cm}
\caption{BTL circuit including connections to the rectifiers, AWG, and dc power supply, as well as the circuit elements required for ac and dc signal decoupling. The flow of the ac (red) and dc (blue) currents along the BTL and through each rectifier is also shown.}
\label{fig:Circuit_AC_DC}
%\vspace{-0.2cm}
\end{figure}

Fig.~\ref{fig:Circuit_AC_DC} shows the BTL and its connection to an arbitrary waveform generator (AWG), modeled with an ideal ac power supply $V_\mathrm{g}$ in series with an impedance $Z_\mathrm{g}$, a dc power supply $V_\mathrm{dc}=W_0$, rectifiers, and some decoupling elements, detailing the current flow of the ac and dc signals. Due to the need for the offset voltage $W_0$, decoupling the ac and dc signals is necessary to prevent short-circuiting and ensure proper current flow along the BTL. To achieve this, a capacitor $C_\mathrm{f} =1$ $\mu$F is connected after the AWG to act as a high-pass filter, while an inductor $L_\mathrm{f} = 680$ $\mu$H is inserted after the dc power supply to serve as a low-pass filter. An additional capacitor $C_\mathrm{f}$ is connected between the end of the BTL and either a short or open circuit termination. The chosen capacitors provide a capacitive reactance $X_{C\mathrm{f}} \approx 0.3$ $\Omega$ at 500 kHz, while the chosen inductor offers an inductive reactance $X_{L\mathrm{f}} \approx 2.1$ k$\Omega$ at the same frequency. This setup ensures that ac signals above 500 kHz (i.e., including all biasing modes) flow through the capacitors but not through the inductor, while signals below 500 kHz (i.e., the dc biasing voltage $W_0$) flow through the inductor but not the capacitors, avoiding short circuits at the opposite end of the BTL.

Since the varactors are connected in reverse bias, the current through them is negligible.
%, meaning the dc current from $V_{\mathrm{dc}}$ mainly flows through the resistances $R_\mathrm{r}$ of the rectifiers when the diodes $D_\mathrm{r}$ are on. On the other hand, the ac current due to the AWG voltage $V_{\mathrm{g}}$ flows mainly through the capacitors $C_\mathrm{r}$, whose impedance has a magnitude of approximately $110$ $\Omega$ at the operating frequency range starting at 7.2 MHz, and thus dominates the parallel path with $R_\mathrm{r}$.
Unlike the dc current, the ac current also reaches the BTL termination through the coupling capacitor $C_\mathrm{f}$. Therefore, the maximum expected dc current flowing from the dc power supply depends on the number of elements $M$, the constant voltage $W_0$ and the resistance $R_\mathrm{r}$. In this case, assuming that the diode $D_\mathrm{r}$ is lossless, the expected dc current is $I_\mathrm{dc} =27(4/10000) = 10.8$ mA.

%DONE***[ELABORATE ON DECOUPLING ELEMENTS, $X_C$, $X_L$, MENTION THE MAX EXPECTED dc CURRENT (AFTER RECTIFIER?) AND DESCRIBE THE FLOW OF BOTH CURRENTS... REF THE FIGURE WITH THE CURRENTS Figure \ref{fig:Circuit_AC_DC}... 

%DONE [mention the via to ground in the layout...]

%DONE [Include CST simulation results showing performance, s-par losses???]

%[DISCUSSION ON THE TD PERFORMANCE OF THE RECTIFIER AND THE Z vs FREQ]

%[Discuss the non-linear behavior of the rectifiers and its correlation with losses (the effect of the dc voltage in the performance (dynamic resistance) of the diode, and the performance considering one single harmonic and multiple harmonics across the rectifier]

%DONE [Include theoretical design, time constant, R and C, features of the diode, ADS simulation results in T-D and also F-D]

%\begin{figure}[htpb]
%\centering
%\includegraphics[width=1\columnwidth]%{Figures/Rectifier_Circuit_v3.png}
%\vspace{-0.5cm}
%\caption{Rectifier circuit. The voltage $v_{\mathrm{in}}=w(x,t)$ is obtained from the BTL while the node of the voltage $v_{\mathrm{out}}=v(x_n)$ is subsequently connected to the CPWG on the RIS board.}
%\label{fig:Rectifier_Circuit_v3}
%\vspace{-0.4cm}
%\end{figure}

%\subsection{Matching Network and Filters}
%- Include theoretical design (Q, BW, cutt-off freq, etc)
%- Include ADS simulation results of the LC network (or any other topology) according to the design (the effect on the whole system (improvement in amplitude of the voltage distribution) can be shown in the Results section)... 

\section{Impact of the Impedances in the dc Pattern}

\subsection{Theoretical Analysis 
and Experimental Results}

%  To attain the full capabilities of this wave-controlled RIS, the maximum voltage measured after a rectifier must be large enough to produce the maximum reflected phase ...
%To fully optimize the reconfiguration capabilities of the RIS, it is essential to ensure that the unit cell achieves the widest possible reflection phase range by maximizing the amplitude of the voltages $w(x_m)$. 
To attain the full capabilities of the wave-controlled RIS, the maximum value of the voltage $w(m)$ measured after the rectifiers must be high enough to enable the widest possible reflection phase range. This voltage amplitude is influenced by proper selection of the frequency $f_\mathrm{b}$ and the BTL termination, either short- or open-circuited. For a short-circuited termination at $x=-L_{\mathrm{left}}$, the input impedance at $x = L+L_{\mathrm{right}}$ 
%for the $n$-th multiple of $f_\mathrm{b}$ 
is given by $Z_{\mathrm{in}} = jZ_0 \mathrm{tan}\left(\kappa\right)$, where $Z_0$ is the characteristic impedance of the microstrip and $\kappa= (\omega_\mathrm{b} n_{\mathrm{slow}}/c)L_{\mathrm{tot}}$. In the particular case where $f_\mathrm{b}=nf_\mathrm{b,0}$, with $n=1,2,3,...$, and $f_\mathrm{b,0}$ is chosen such that the wavelength $\lambda_{\mathrm{b,0}} = c/(n_{\mathrm{slow}} f_{\mathrm{b,0}}) = 4 L_{\mathrm{tot}}$, the input impedance is $Z_{\mathrm{in},n} = jZ_0\mathrm{tan}\left(n\pi/2\right)$. Under this condition, $Z_{\mathrm{in},n}$ becomes infinite for odd multiples of $f_\mathrm{b,0}$ (i.e., $n=1,3,5,...$), and zero for even multiples (i.e., $n=2,4,6,...$). As a result, the voltage at $x = L+L_{\mathrm{right}}$ reaches a maximum for odd multiples and a minimum for even multiples of $f_\mathrm{b,0}$. 

\begin{figure}[htpb]
%\vspace{-0.3cm}
\centering
\includegraphics[width=0.88\columnwidth]{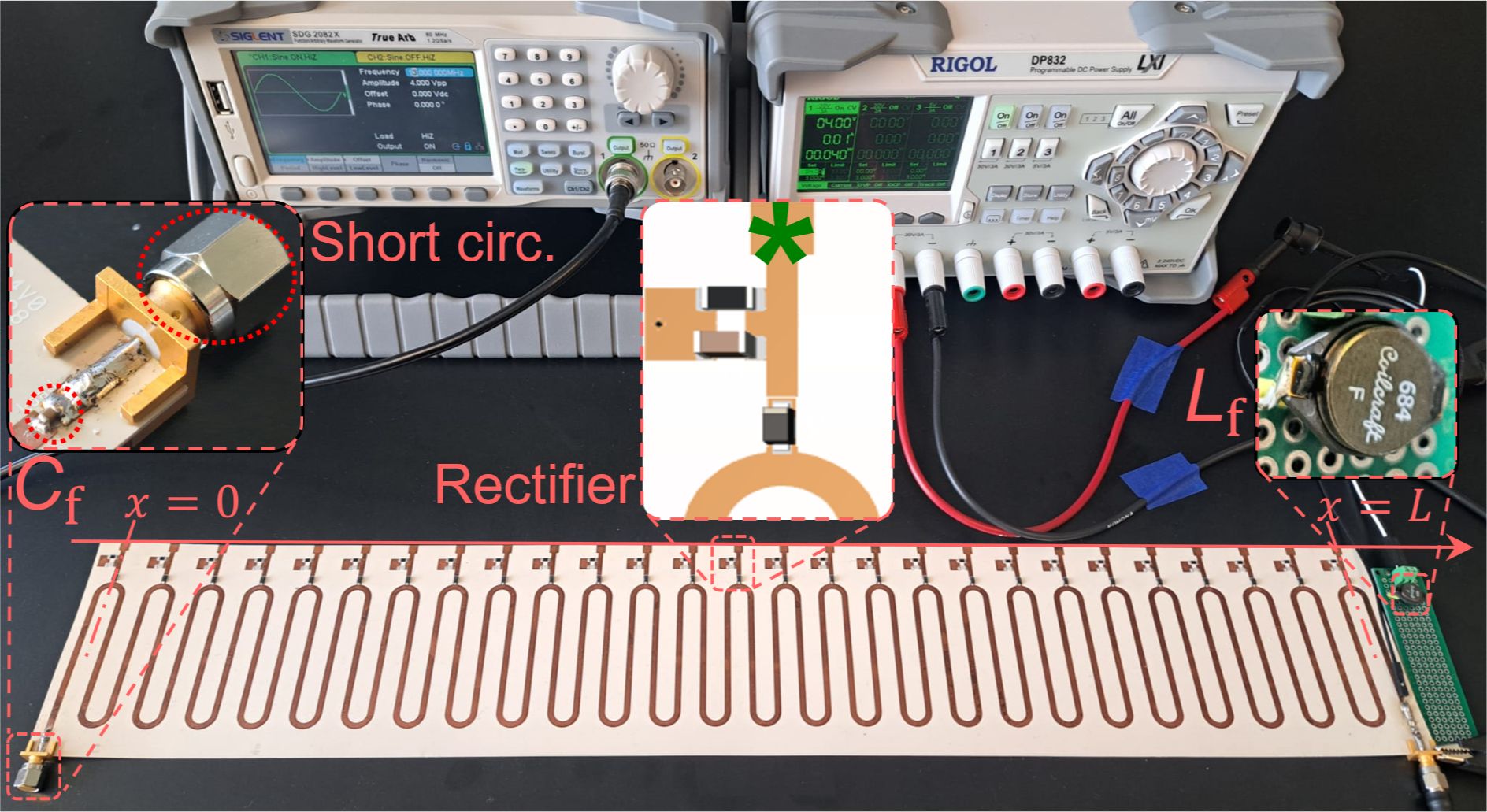}
%\vspace{-0.35cm}
\caption{Printed meandered BTL backed by the grounded dielectric substrate RO3010 connected to the Siglent SDG6052X AWG, and the RIGOL DP832 dc power supply.}
\label{fig:Experiment_setup}
%\vspace{-0.2cm}
\end{figure}

\begin{figure}[htpb]
%\vspace{-0.3cm}
\centering
\includegraphics[width=0.9\columnwidth]{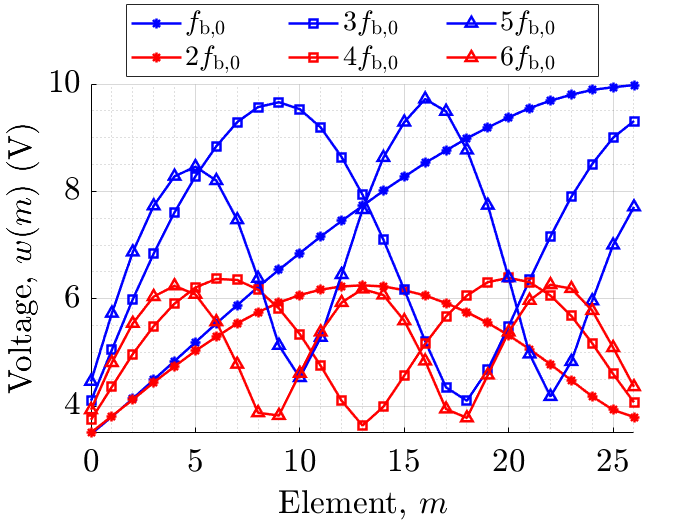}
%\vspace{-0.35cm}
\caption{Measured dc bias voltage $w(m)$ along the BTL at positions $x_m = md_x$ for six individual BSWs, with the BTL terminated by a short circuit at $x=-L_{\mathrm{left}}$. The red and blue curves represent cases where $f_\mathrm{b}$ is an even or odd multiple of $f_\mathrm{b,0}$, respectively.}
\label{fig:Comparison_SC}
%\vspace{-0.2cm}
\end{figure}

In order to experimentally compare the voltage distribution along the BTL for even and odd multiples of $f_\mathrm{b,0}$, we use the experimental setup shown in Fig.~\ref{fig:Experiment_setup}, which includes an AWG, power supply, decoupling elements, and BTL. For the BTL design presented here, $f_\mathrm{b,0} = 7.18$ MHz. After setting $V_\textrm{g}=10$ V, Fig.~\ref{fig:Comparison_SC} shows the voltages $w(m)$ measured with a multimeter at positions $x_m = md_x$, $m = 1,2,..., M-1$, after each rectifier along the BTL terminated in a short circuit, for six BSWs with frequencies that are multiples of $f_\mathrm{b,0}$. The results reveal a maximum amplitude difference exceeding $3$ V between odd and even multiples of $f_\mathrm{b,0}$. However, the minimum values remain close to $W_0$ in all cases, slightly falling below $W_0$ due to losses associated with the rectifiers. The voltages at the first and last elements show slight deviations from the expected values, with these deviations increasing at higher frequencies since the measurements were taken half a period $(d_x/2)$ before the ends of the BTL, corresponding to the lengths of $L_\mathrm{left}$ and $L_\mathrm{right}$.

\begin{figure}[htpb]
%\vspace{-0.3cm}
\centering
\includegraphics[width=0.9\columnwidth]{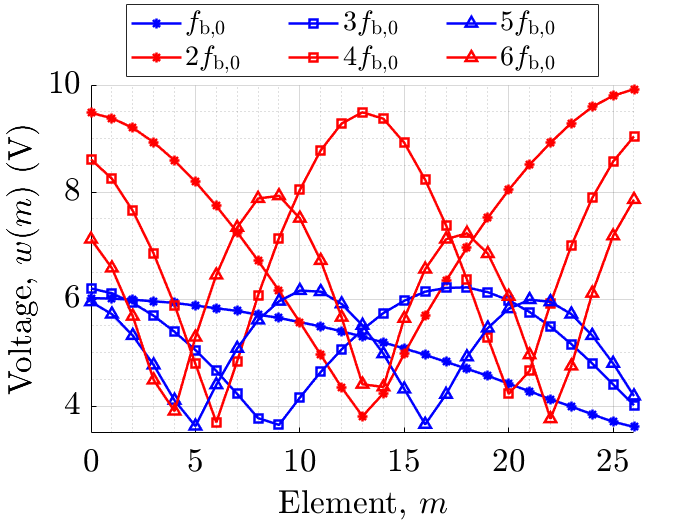}
%\vspace{-0.35cm}
\caption{Similar to Fig.~\ref{fig:Comparison_SC}, but with the BTL terminated by an open circuit at $x=-L_{\mathrm{left}}$.}
\label{fig:Comparison_OC}
%\vspace{-0.2cm}
\end{figure}

Similarly, for the case where the BTL is terminated in an open circuit at $x=-L_{\mathrm{left}}$, the input impedance at $x=L+L_{\mathrm{right}}$ 
%for a frequency multiple $nf_\mathrm{b}$ 
is $Z_{\mathrm{in}} = jZ_0 \mathrm{cot}\left(\kappa\right)$. By choosing $f_\mathrm{b}=nf_\mathrm{b,0}$ as before, the input impedance for $n$-multiples of $f_\mathrm{b,0}$ becomes $Z_{\mathrm{in},n} = jZ_0\mathrm{cot}\left(n \pi/2\right)$. This causes $Z_{\mathrm{in},n}$ to become infinite for even multiples of $f_\mathrm{b,0}$ and zero for odd multiples, resulting in a maximum voltage at $x = L+L_{\mathrm{right}}$ for odd multiples and a minimum for even multiples of $f_\mathrm{b,0}$. Fig.~\ref{fig:Comparison_OC} presents a comparison of the voltages $w(m)$ measured along the BTL with an open-circuit termination. Similar to previous cases, the results reveal a maximum amplitude difference of over $3$ V between odd and even multiples of $f_\mathrm{b,0}$, with higher amplitudes for even multiples in this case. As before, the minimum values remain close to $W_0$ for all cases, while slight deviations are observed in the first and last elements due to the $d_x/2$ shift before the BTL endpoints.

\begin{figure}[htpb]
\centering
\includegraphics[width=0.8\columnwidth]{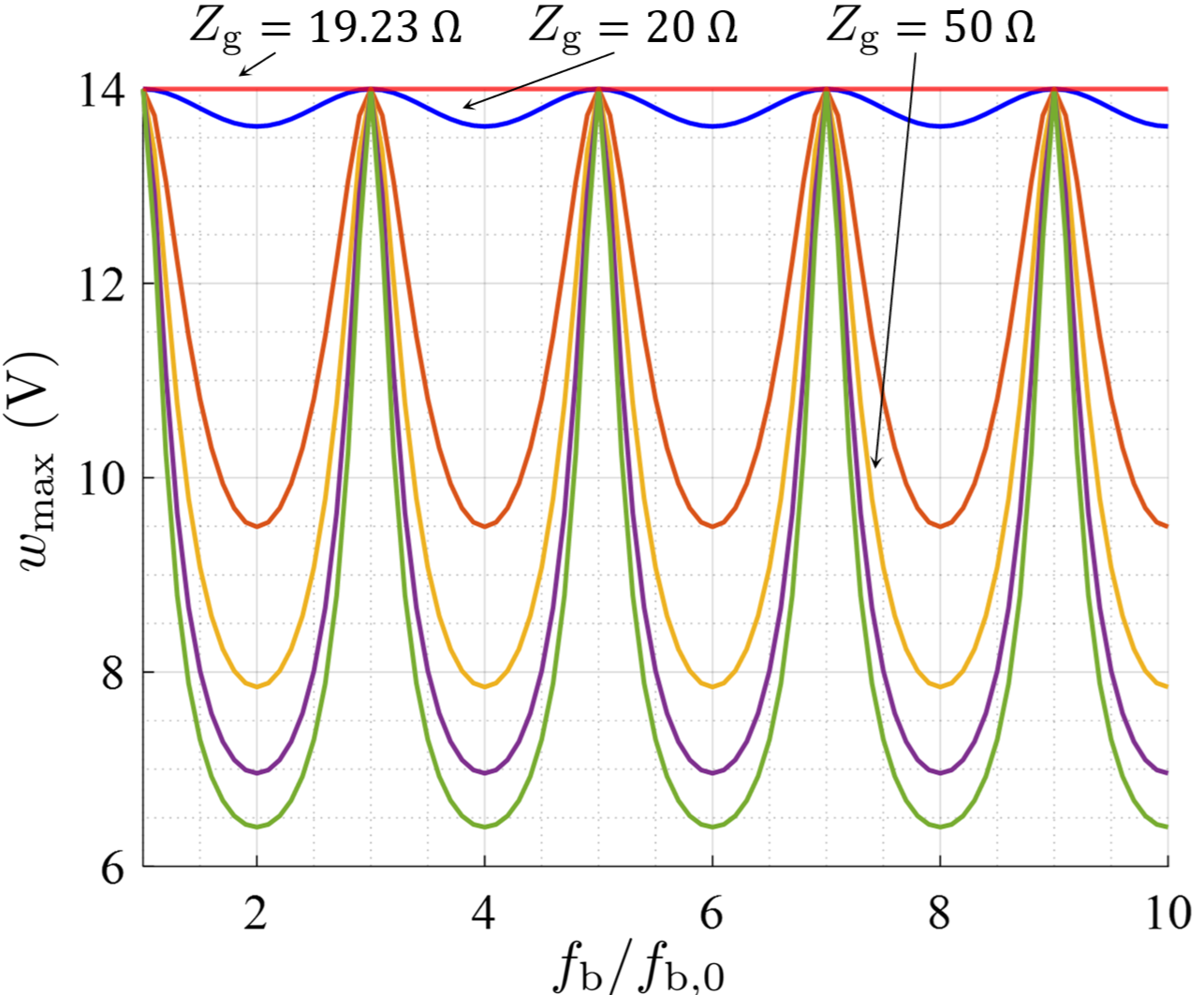}
%\vspace{-0.5cm}
\caption{Maximum dc biasing voltage along the BTL for different frequencies for a short-circuit BTL termination, assuming $V_\mathrm{g}=10$ V and $Z_{\mathrm{g}}$ between $Z_0$ (red curve) and $100$ $\Omega$ (green curve). When $Z_{\mathrm{g}} = Z_0$, the maximum voltage is the same for even and odd multiples of $f_\mathrm{b}$.}
\label{fig:A_vs_SclFactor}
%\vspace{-0.4cm}
\end{figure}

The difference observed in the amplitude for open- and short-circuited terminations is attributed to the effect of the output impedance $Z_{\mathrm{g}}$ of the AWG that injects the periodic time-domain signal at $nf_\mathrm{b,0}$. As an example, Fig.~\ref{fig:A_vs_SclFactor} shows the maximum dc biasing voltage $w_\mathrm{max}=\max\limits_{m}\ w(m)=W_\mathrm{b}+W_{\mathrm{0}}$ versus the frequency $f_\mathrm{b}$ for a few values of the AWG impedance $Z_{\mathrm{g}}$ from $Z_0$ ($19.23$ $\Omega$, red curve) to $100$ $\Omega$ (green curve), assuming a short-circuit termination. These results are obtained using an analytical model of the BTL. 

%In this analysis, the scaling factor $a$ is introduced to adjust the frequency continuously, unlike the integer index $n$, allowing the observation of frequencies between integer multiples. When $a=1$, the frequency $f_\mathrm{b}$ corresponds to the condition where $\lambda_{\mathrm{b}} = 4L_{\mathrm{tot}}$.

We observe that for frequencies that are odd multiples of $f_\mathrm{b,0}$, $w_\mathrm{max}$ remains constant and reaches its largest possible value. On the contrary, for frequencies that are even multiples of $f_\mathrm{b,0}$, $w_\mathrm{max}$ varies depending on the impedance of the AWG. This behavior arises because the relationship $W_\mathrm{b}=(Z_0/Z_{\mathrm{g}})V_{\mathrm{g}}$ applies for even multiples of $f_\mathrm{b,0}$, while for odd multiples we have $W_b=V_\mathrm{g}$. Consequently, to maximize $w_\mathrm{max}$ for {\em both} even and odd multiples of $f_\mathrm{b,0}$, the BTL should be designed with a characteristic impedance $Z_0$ that matches the AWG impedance $Z_{\mathrm{g}}$. The mathematical derivations supporting this analysis are provided in Appendix~B.%\ref{Appendix:B}.

In summary, from a practical perspective, two different design criteria can be employed to maximize $w_\mathrm{max}$ and ensure that it remains constant for all frequencies. The first approach involves selecting $f_\mathrm{b} = nf_\mathrm{b,0}$ with $n$ odd, which guarantees maximization regardless of the relationship between $Z_\mathrm{g}$ and $Z_0$. For the second approach, the BTL is designed with $Z_0 = Z_\mathrm{g}$, which ensures maximization of $w(m)$ for every value of $f_\mathrm{b}$. The second approach offers an additional advantage when multiple BSWs are used \cite{Ben-Itzhak24}, as it effectively doubles the number of biasing modes within the same frequency range while also making $f_\mathrm{b}$ independent of the total length $L_{\mathrm{tot}}$.

%In the end, selecting odd multiples of $f_\mathrm{b,0}$ when $\lambda_\mathrm{b}=4L_\mathrm{tot}$ ensures the maximization of $w(x_m)$ regardless of the relationship between $Z_\mathrm{g}$ and $Z_0$. However, designing the BTL with $Z_0=Z_\mathrm{g}$ allows for the maximization of $w(x_m)$ when using all the multiples of $f_\mathrm{b}$, independent of the value of $\lambda_\mathrm{b}$ and, therefore, duplicating the number of biasing modes within the same frequency range.

% Amplitude of the voltage A [CHANGE TO $W_\mathrm{b}$] 

%\section{Experimental Results and Discussion}
%\subsection{Frequency Response}
%- Include results with the VNA...
%\subsection{Time-domine response}
%- Description of the process of the generation of harmonics by using the MATLAB script...
%- Include results with AWG-oscilloscope-multimeter...

\subsection{Comparison with Numerical Model}

\begin{figure}[htpb]
%\vspace{-0.3cm}
\centering
\includegraphics[width=1\columnwidth]{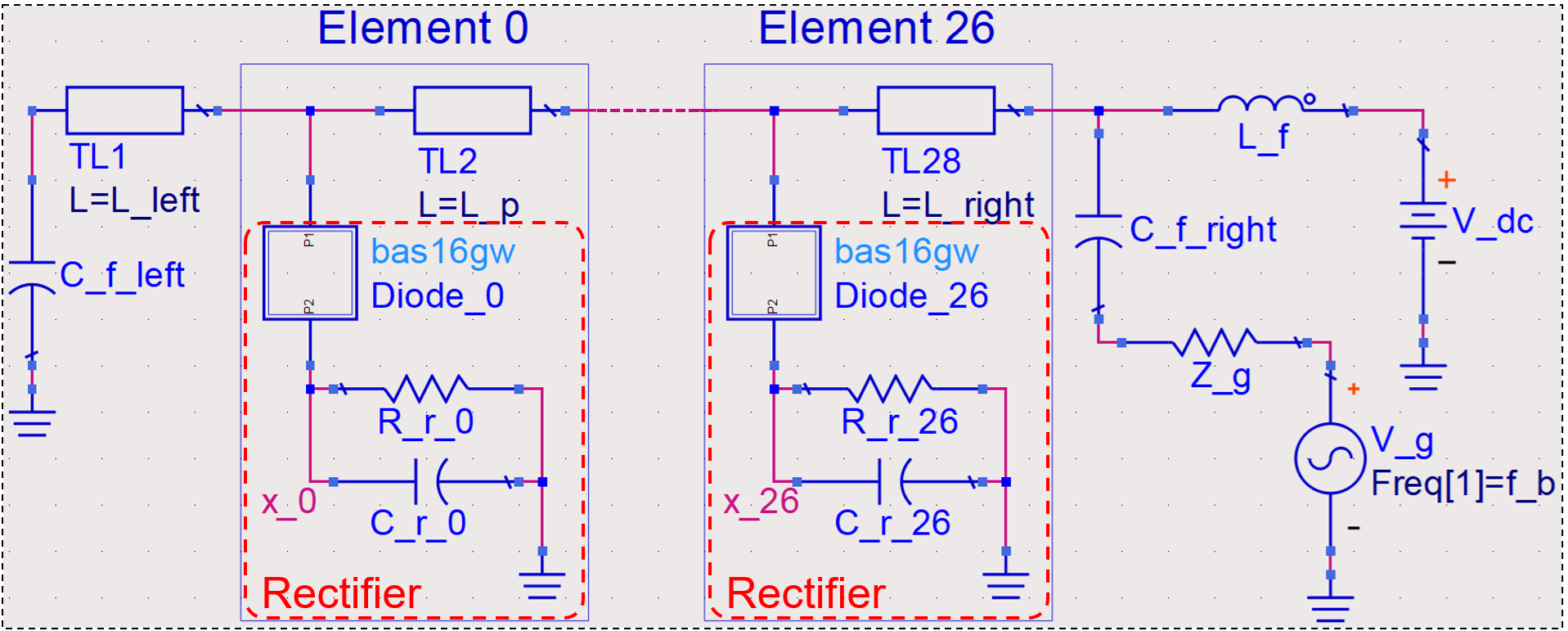}
%\vspace{-0.35cm}
\caption{ADS simulation setup: The ac and dc voltage generators and decoupling elements are connected to the rightmost cell. All unit cells are arranged in a cascaded configuration and include a rectifier circuit. On the left side, the cascade is terminated in a short circuit ($\approx 1/(j \omega C_\mathrm{f})$) after a TL with length $L_\mathrm{left}$. Each unit cell is modeled as a transmission line segment with characteristic impedance $Z_0$ and length $L_\mathrm{p}$, except the rightmost cell that has length $L_\mathrm{right}$.}
\label{fig:ADS_BTL_setup}
%\vspace{-0.2cm}
\end{figure}

In addition to the analytical model, a numerical implementation of the BTL is developed in ADS to extend the analysis. The simulation setup shown in Fig.~\ref{fig:ADS_BTL_setup} includes an AWG modeled as an ideal ac source of amplitude $V_\mathrm{g}=10$ V in series with an impedance of $Z_\mathrm{g} = 50$ $\Omega$, an ideal dc power source, and the required decoupling elements. These components are connected to the rightmost unit cell of the BTL after a transmission line segment of length $L_\mathrm{right}$ with characteristic impedance $Z_0$. Each unit cell has a rectifier circuit composed of a diode ($D_\mathrm{r}$), a resistor ($R_\mathrm{r}$), and a capacitor ($C_\mathrm{r}$). The diode is modeled by importing the touchstone file provided by the vendor. Unit cells are cascaded and modeled as identical transmission lines of length $L_\mathrm{p}$, each with the same rectification circuit. On the left side, the  
%last 
unit cell is connected to a transmission line segment of length $L_\mathrm{left}$, followed by the capacitor ``C\_f\_left'' that is either connected to ground, modeling a short-circuit termination as shown in the figure, or left floating, modeling an open-circuit termination. In this ADS configuration, both decoupling capacitors, labeled ``C\_f\_left'' and ``C\_f\_right'', are assigned the same value $C_\mathrm{f}$.
 
%\subsection{Single Tones along 27 Elements}

\begin{figure}[htpb]
\centering
\includegraphics[width=0.9\columnwidth]{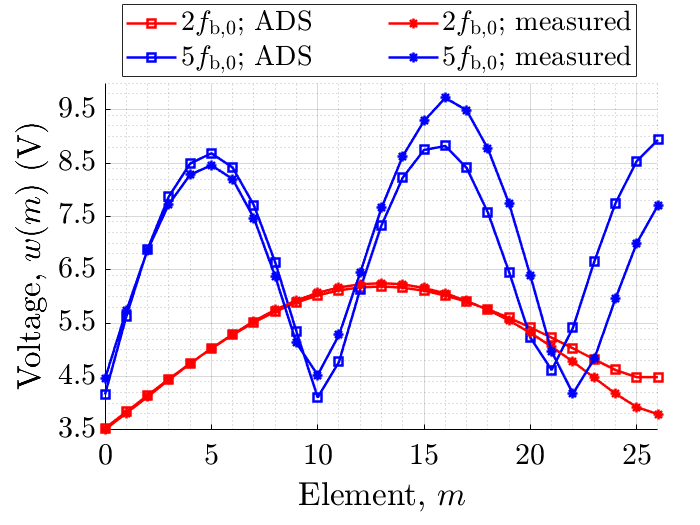}
%\vspace{-0.5cm}
\caption{Voltage distribution along the BTL for two different single tones with frequencies $2f_\mathrm{b,0}$ and $5f_\mathrm{b,0}$ where $W_\mathrm{b}=10$ V and the dc component is $W_\mathrm{0}=4$ V.}
\label{fig:ADS_Exp_Volt_Distr_N27}
%\vspace{-0.4cm}
\end{figure}

Fig.~\ref{fig:ADS_Exp_Volt_Distr_N27} 
compares the voltage $w(m)$ evaluated using the ADS model with that from experimental results for two different BSW frequencies: one using an even multiple of $f_\mathrm{b,0}$ and the other using an odd multiple, both with a short-circuit termination. The results show good agreement between simulation and measurement in the case of the even multiple of $f_\mathrm{b,0}$, where $w_\mathrm{max}$
%the voltage distribution amplitude 
is minimized. 
%The discrepancy between simulation and measurement is higher for the odd multiple case, where $w_\mathrm{max}$
%%the amplitude 
%is maximized. This difference arises because for even multiples, $Z_{\mathrm{in}}$ approaches 0 $\Omega$, while for the odd multiple cases, $Z_{\mathrm{in}}$ becomes large and the numerical limitations in ADS restrict the accuracy of simulating such high impedance values, leading to a greater deviation from the measured data.
The discrepancy between simulation and measurement increases with distance from the short-circuit termination. This can be attributed to the fact that the TL segments that model the BTL in ADS are straight and therefore do not account for geometric effects that are neglected in the model, such as the curvature present in the meandered line design, as well as coupling effects between parallel segments.

\begin{figure}[htpb]
\centering
\includegraphics[width=0.9\columnwidth]{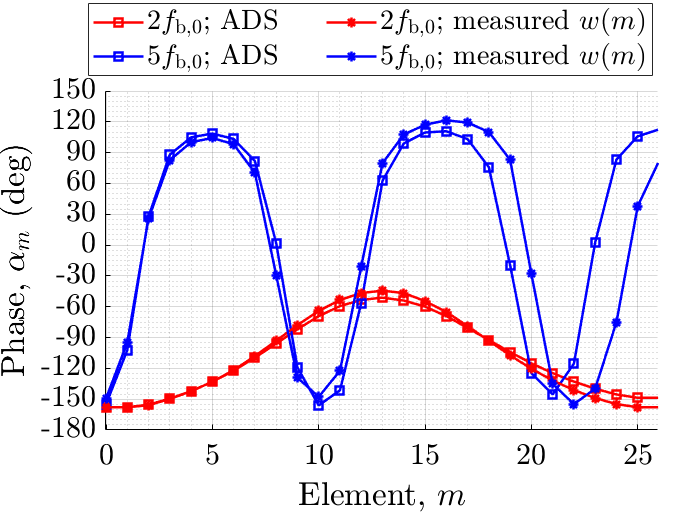}
%\vspace{-0.5cm}
\caption{Reflection phase along the linear RIS array for single tones with frequencies $2f_\mathrm{b,0}$ and $5f_\mathrm{b,0}$.}
\label{fig:ADS_Exp_Phase_Distr_N27}
%\vspace{-0.4cm}
\end{figure}

\begin{figure}[htpb]
\centering
\includegraphics[width=0.9\columnwidth]{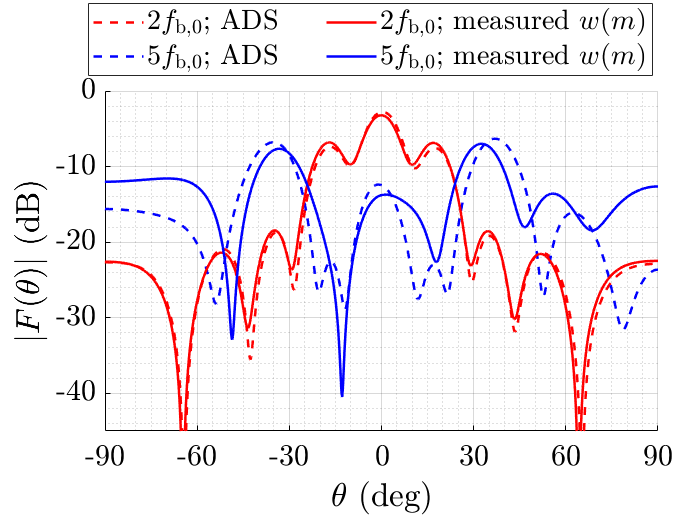}
%\vspace{-0.5cm}
\caption{Reflected radiation pattern generated by the linear RIS array assuming normal plane wave incidence for single-tone frequencies $2f_\mathrm{b,0}$ and $5f_\mathrm{b,0}$.}
\label{fig:ADS_Exp_Rad_Pattern_N27}
%\vspace{-0.4cm}
\end{figure}

Fig.~\ref{fig:ADS_Exp_Phase_Distr_N27} compares the reflection-phase distribution along the RIS, determined using the relationship in Fig.~\ref{fig:Phase_Mag_vs_V_freq}(\subref*{fig:Phase_Mag_vs_voltage}) using both ADS and measured voltages. As expected, the case with the even multiple $2f_\mathrm{b,0}$ exhibits good agreement between simulation and measurement, although the phase variation is limited, ranging from approximately $-150^\circ$ to $-45^\circ$ due to the relatively narrow range of biasing voltages. In contrast, the odd multiple case, $5f_\mathrm{b,0}$, shows a much broader phase variation, ranging from about $-150^\circ$ to $120^\circ$, but with a more noticeable mismatch between ADS and experimental results for large $m$. Finally, Fig.~\ref{fig:ADS_Exp_Rad_Pattern_N27} illustrates the expected scattered radiation patterns calculated using the field reflection magnitude and phase for each RIS element obtained from both the simulated and measured cases. The scattered pattern is evaluated using the normalized array factor, defined as
\begin{equation}
\label{eq:NormAF}
F\left(\theta\right)=\frac{1}{M}\sum_{m=0}^{M-1}R_{m}e^{jmkd_x\sin\left(\theta\right)+j\alpha_{m}},
\end{equation}
where $\alpha_m$ and $R_m$, previously shown in Fig.~\ref{fig:Phase_Mag_vs_V_freq}(\subref*{fig:Phase_Mag_vs_voltage}) , represent the phase and magnitude of the reflection coefficient for the $m$-th unit cell given the applied voltage $w(m)$ as determined by~(\ref{eq:BTL_Volt_Retif_Single}). For the two selected values $2f_\mathrm{b,0}$ and $5f_\mathrm{b,0}$, the resulting radiation patterns concentrate the reflected energy near $\theta = 0^\circ$ and $\theta = \pm 32^\circ$.
%Since arbitrary frequencies were chosen, the radiation patterns do not concentrate reflection in a specific direction. 
However, the key takeaway is the comparison between the two cases, where, as expected, the even multiple of $f_\mathrm{b,0}$ yields better agreement.

\section{Reflective Control Using a Single Biasing Frequency}
\begin{figure}[htpb]
\centering
\includegraphics[width=1\columnwidth]{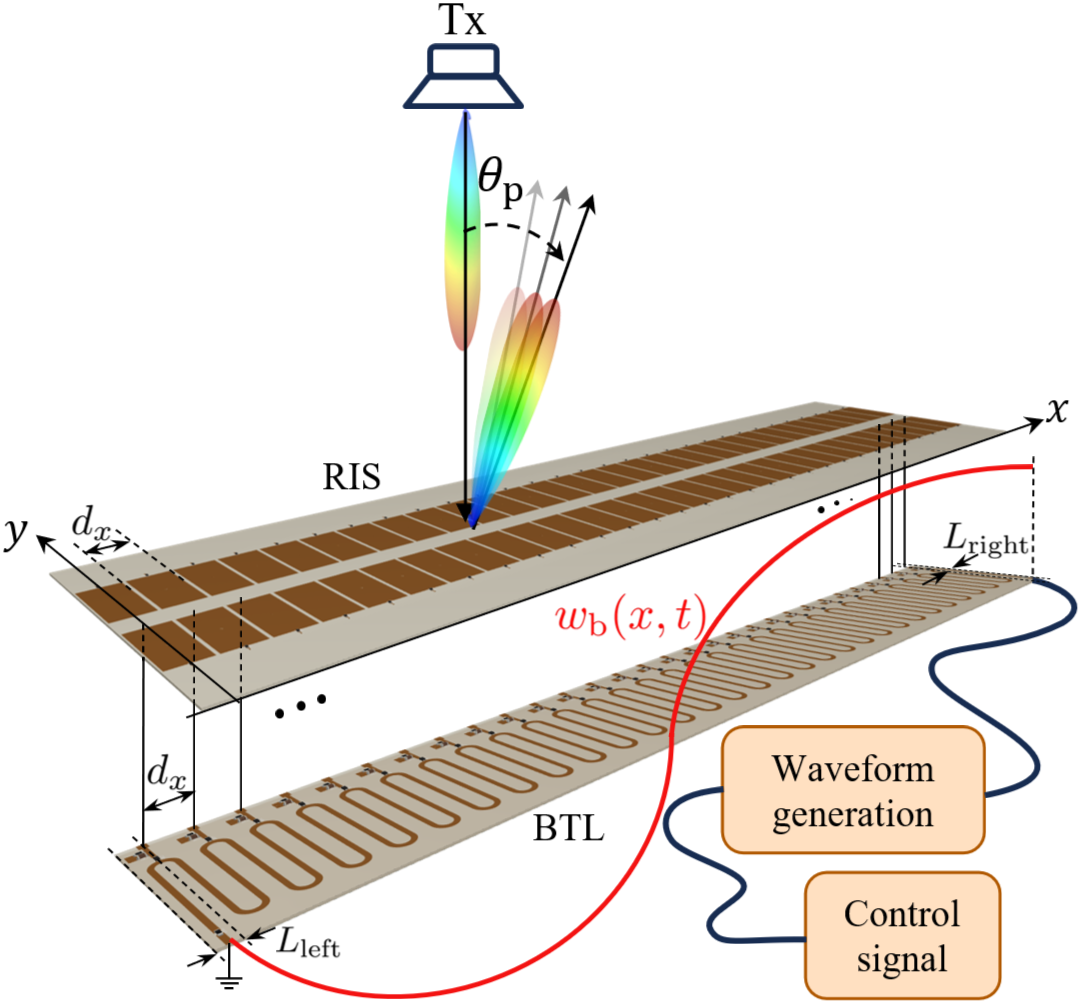}
%\vspace{-0.5cm}
\caption{Wave-controlled RIS consisting of two boards, as described in \cite{Ayanoglu22}. The top layer contains $M$ pairs of RIS elements arranged in a row along the $x$ direction, each connected to a varactor diode. In the bottom layer, a single BSW propagates along the BTL, generating the biasing voltages $w(m)$, which are sampled at positions spaced by $d_x$ in the $x$ direction.}
\label{fig:Rad_Pattern_N27}
%\vspace{-0.4cm}
\end{figure}

We study the beam- and null-steering capabilities of the wave-controlled RIS with a {\em single biasing frequency}, as visualized in Fig.~\ref{fig:Rad_Pattern_N27}. Although using a single biasing frequency does not allow for the more flexible reflective control permitted by multiple BSWs along the transmission line \cite{Ben-Itzhak24}, it will be shown that this strategy can be used for reflective control in the following ways: first for creating nulls at broadside which can, for example, enable designs to minimize the radar cross section, and second for single beam steering, with limitations at large angles. Further possibilities for extending this control will also be discussed.

The radiation pattern plots in the analysis below result from the proposed design assuming the BTL is lossless and impedance matched to the AWG, but the analysis applies in more general settings as well. These assumptions allow (\ref{eq:BTL_Volt_Retif_Single}) to be used to determine the exact voltage seen at different points along the BTL. It is important to note that when the BTL is impedance matched to the AWG, the length of the BTL to the right of the last unit cell, $L_\mathrm{right}$, has no impact on the voltage pattern seen by the element rectifiers. %along the BTL's length.

\subsection{Nulling the Specular Reflection}

For many applications, including those described in \cite{Cheng_RCS_Metasurf_15} and \cite{Xu_Radar_Dyn_Jamming_21}, it is necessary to reduce the specular reflection of the incident radiation. 
Maximizing the reflection towards a different direction does not necessarily fully diminish the specular reflection, due to the non-perfect phase gradient along the RIS. The wave-controlled RIS with multiple BSWs described in \cite{Ben-Itzhak24} was able to suppress the specular component by explicitly including it as a constraint in its optimization criteria. The wave-controlled RIS with a single biasing frequency can also reduce the specular reflection, but with more limited degrees of freedom.

\begin{figure}[t]
\label{Design}
\begin{centering}
\includegraphics[width=1\columnwidth]{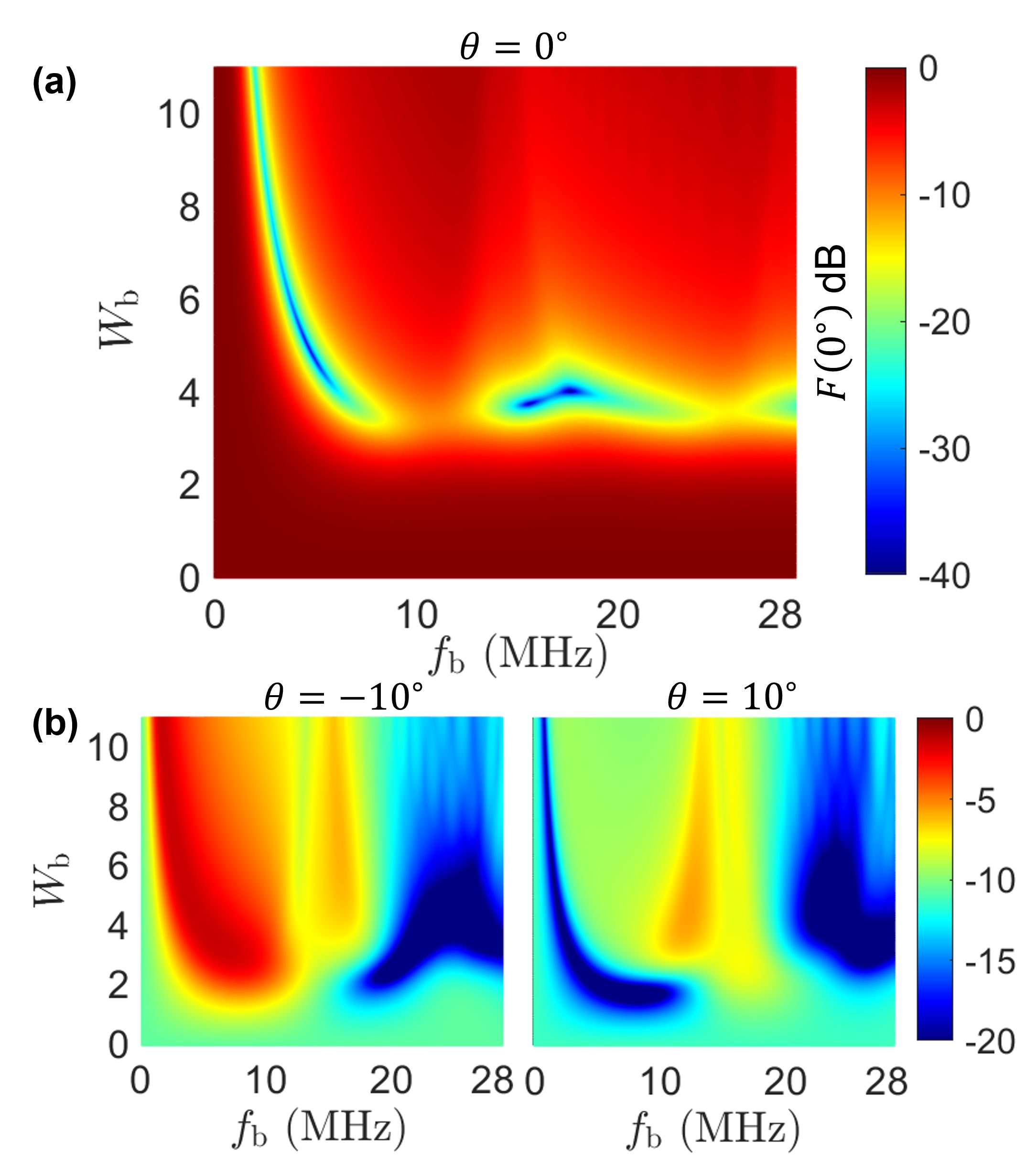}
\par\end{centering}
\caption{Colormap of normalized RIS array factor plotted against BSW frequency $f_\mathrm{b}$ and amplitude $W_\mathrm{b}$, with the BTL terminated in a short circuit. We assume normal incidence. (a) Radiation at $\theta=0^\circ$: minimal specular reflections occurs for frequencies $f_\mathrm{b}$ greater than 7 MHz around $W_\mathrm{b} = 4$ V. (b) Radiation at $-10^\circ$ is maximized for frequencies below 10 MHz. The colormap for $10^\circ$ shows no symmetrical ``ghost'' beam when the beam is directed towards $-10^\circ$.} \label{fig:Colormaps}
\end{figure}

Fig.~\ref{fig:Colormaps}(a) shows the RIS reflection at $\theta=0^\circ$ for the above design due to normal incident illumination against the biasing frequency $f_\mathrm{b}$ and the wave amplitude $W_\mathrm{b}$ in (\ref{eq:BTL_Volt_Retif_Single}). The dc bias, $W_0$, is not considered to be a tunable parameter here and is set to 4 V as this is the minimum biasing voltage required for the varactor specified in this design (see Table \ref{tab:Cv_Rv_Varactor}), and increasing the dc bias will reduce the range of controllable phases. 
From this colormap, it can be seen that for most combinations of frequency and wave amplitude, the specular reflection at $\theta=0^\circ$ is non-negligible. However, there are combinations of these parameters that result in significantly reduced specular reflection. The region of reduced reflection begins at 2 MHz with $W_\mathrm{b}=12$ V, and is approximately constant at $W_\mathrm{b} = 4$ V for frequencies larger than 7 MHz, although the magnitude fluctuates with frequency. The minimal reflections occurring for $W_\mathrm{b}=4$ V coincide with the approximately linear region of the voltage-to-phase relationship shown in Fig.~\ref{fig:Phase_Mag_vs_V_freq}(\subref*{fig:Phase_Mag_vs_voltage}).

The region of minimal specular reflection results in the radiation being directed elsewhere, as seen by comparing Figs.~\ref{fig:Colormaps}(a) and~\ref{fig:Colormaps}(b) where we observe radiation at $\theta=\pm 10^\circ$. Fig.~\ref{fig:Colormaps}(b) shows that the redirected radiation at $\theta=-10^\circ$ is maximized for frequencies lower than $10$ MHz without a symmetric peak at $\theta=10^\circ$; the symmetric peak often arises with beams steered to larger angles \cite{Bradshaw_MultiBoard_2025}. 

\subsection{Single Beam Control Near Broadside}
For single beam control, the ideal phase difference between adjacent RIS elements is 
\begin{equation}\label{eq:SingleBeam}
    \alpha_{m+1}-\alpha_m = -2\pi d_x \sin(\theta_{\mathrm{p}})/\lambda_\mathrm{c},
\end{equation}
where $\theta_{\mathrm{p}}$ is the desired beam angle (peak radiation direction) and $\lambda_\mathrm{c}$ is the free-space wavelength of the carrier \cite{Balanis_Antenna_15}. Hence, the phase for the element at $x_m = md_x$ on the BTL is 
\begin{equation}\label{eq:SingleBeamElementPhase}
\alpha_m = \alpha_0 -2\pi x_m \sin(\theta_{\mathrm{p}})/\lambda_\mathrm{c}. 
\end{equation}
For large angles, achieving this phase gradient across the RIS leads to phase wrapping. However, for small angles where $|\alpha_{M-1}-\alpha_0|<360^\circ$ which do not lead to phase wrapping, this linear phase response along the RIS can be created by a single BSW. This method works best when $L_\mathrm{left}$ is small as it allows for maximum phase variation across the elements. This also follows from the discussion in Sec. IV as the length $L_\mathrm{right}$ does not impact the voltage pattern seen by element $M-1$ when the BTL is impedance matched to the AWG.

\begin{figure}[t]
\label{Design}
\begin{centering}
\includegraphics[width=1\columnwidth]{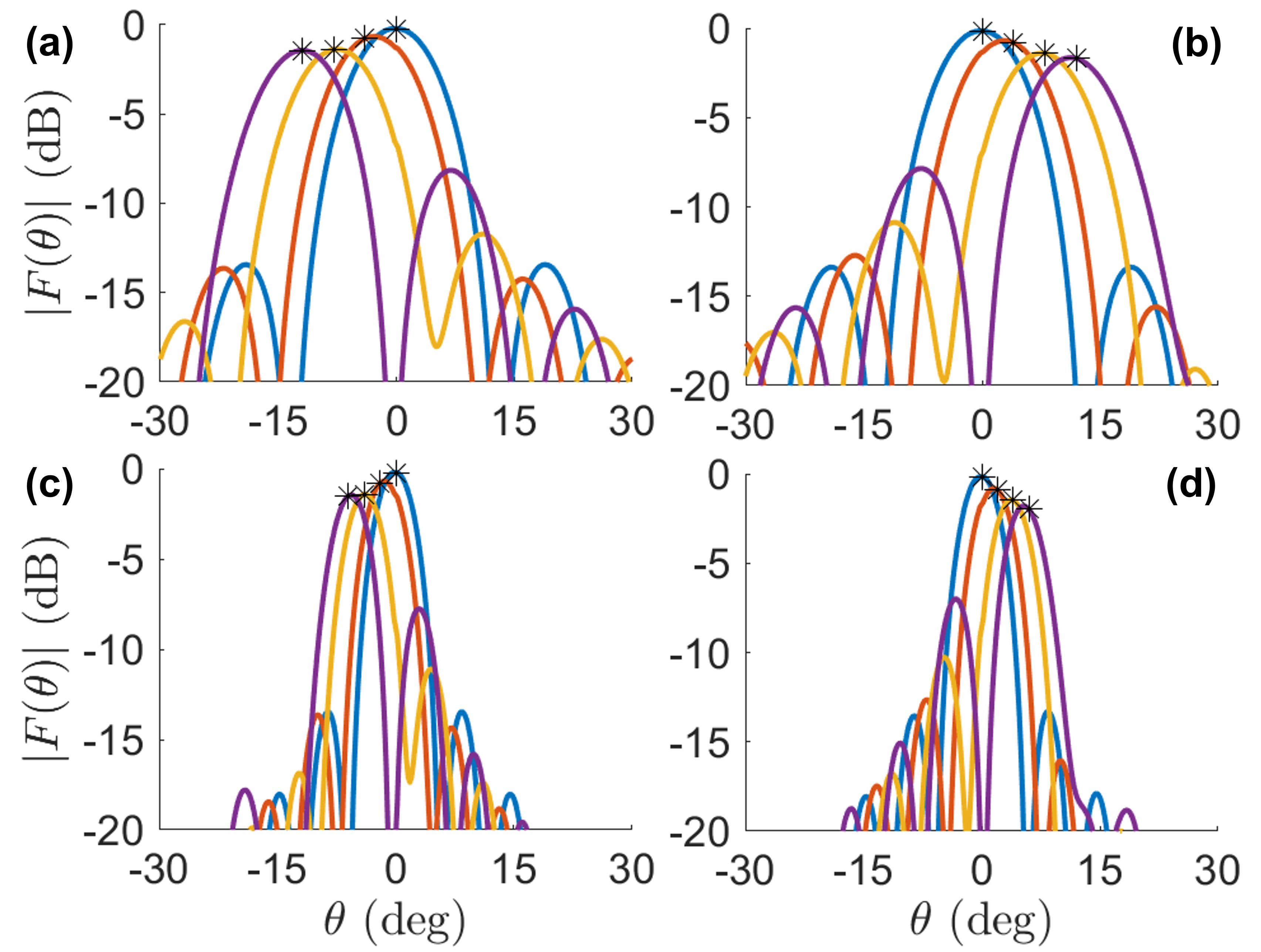}
\par\end{centering}
\caption{Normalized radiation patterns at $f_\mathrm{c} = 2.45\ \mathrm{GHz}$ for the wave-controlled RIS, obtained by optimizing $W_\mathrm{b}$ and $f_\mathrm{b}$ to maximize the radiated power at the angles marked by an asterisk; (a) and (b) correspond respectively to short- and open-circuit BTL terminations, with 27 RIS elements, with $\theta_{\mathrm{p}}=$ $0^\circ, \pm4^\circ, \pm8^\circ,$ and $\pm12^\circ$; (c) and (d) correspond to short- and open-circuited BTLs, with 60 RIS elements, with with $\theta_{\mathrm{p}}=$ $0^\circ, \pm2^\circ, \pm4^\circ,$ and $\pm6^\circ$.} 
\label{fig:SmallAngleRadPatterns}
\end{figure}

There are two aspects of this design that limit the linear phase gradient from being exactly reproduced across the board. The first is the nonlinearity of the voltage-to-phase relationship as seen in Fig.~\ref{fig:Phase_Mag_vs_V_freq}(\subref*{fig:Phase_Mag_vs_voltage}). This problem can be addressed by using a wideband unit cell design, which would have a more linear voltage-to-phase relationship, such as the dual resonant element presented in \cite{LinearElement}. The second factor is the sinusoidal nature of the voltage gradient across $L$. For the short-circuit termination, the voltage gradient is approximately linear when $(2\pi f_\mathrm{b} n_{\mathrm{slow}}(x_{M-1}+L_\mathrm{left}))/c \ll 1$, allowing the small-angle approximation to be applied to (\ref{eq:BTL_Volt_Retif_Single}). Beyond this limit and for either termination, linearity diminishes but the phase gradient is still monotonically increasing or decreasing along the BTL while $\lambda_{\mathrm{b}} > 4(L+L_\mathrm{left})$. This condition still allows for a highly directive beam with relatively low sidelobes and a minor loss of power at progressively larger angles, as seen in Fig.~\ref{fig:SmallAngleRadPatterns}. The wavelength $\lambda_{\mathrm{b}} > 4(L+L_\mathrm{left})$ will limit the maximum allowed $\theta_\mathrm{p}$ as it corresponds to the case where $|\alpha_{M-1}-\alpha_0|<360^\circ$. Beyond this limit, when $\lambda_{\mathrm{b}} < 4(L+L_\mathrm{left})$, the phase gradient is no longer monotonic leading to the generation of other radiation peaks specifically in the $-\theta_{\mathrm{p}}$ direction.

Due to these limitations, a trade-off exists between the length of the array, $L$, and the maximum usable angle for beam steering, $\theta_\mathrm{p}$. Larger arrays will have smaller beamwidths but will be more limited in beam steering angles. This is shown in Fig.~\ref{fig:SmallAngleRadPatterns} which illustrates two beam steering examples for both the open- and short-circuit terminations for a wave-controlled RIS with either 27 or 60 elements. At the specified angles, optimal $f_\mathrm{b}$ and $W_\mathrm{b}$ are obtained through a search of the parameter space (see Table \ref{tab:Beamsteering}), while $\theta_p=0^\circ$ is maximized when $W_\mathrm{b}=0$ for any frequency. All of the patterns shown in the figure maintain the $|\alpha_{M-1} - \alpha_0| < 360^\circ$ requirement since the radiation patterns for larger $\theta_\mathrm{p}$ would have rapidly decreasing directivity in the desired direction.

\begin{table}[]
\caption{Values of $f_\mathrm{b}$ and $W_\mathrm{b}$ for the curves in Fig. \ref{fig:SmallAngleRadPatterns}.}
\centering
\begin{tabular}{|c|c|c|c|c|c|c|c|}
\hline
\multicolumn{4}{|c|}{Short-circuit termination} & \multicolumn{4}{|c|}{Open-circuit termination} \\ \hline
& $\theta_\mathrm{p}$  & $f_\mathrm{b}$ (MHz) & $W_\mathrm{b}$ &
& $\theta_\mathrm{p}$  & $f_\mathrm{b}$ (MHz) & $W_\mathrm{b}$ \\ \hline
\multirow{3}{*}{(a)} & -4$^\circ$ & 1.2 & 7.3 & \multirow{3}{*}{(b)} &  4$^\circ$  & 8.1 & 1.8 \\ %\hline
                       & -8$^\circ$ & 6.0 & 2.9 &                        & 8$^\circ$  & 7.5 & 2.7 \\ %\hline
                       & -12$^\circ$ & 2.0 & 10.8 &                       & 12$^\circ$  & 7.5 & 3.7 \\ \hline
\multirow{3}{*}{(c)} & -2$^\circ$ & 0.7 & 6.2 & \multirow{3}{*}{(d)} & 2$^\circ$ & 3.6 & 1.9  \\ %\hline
                       & -4$^\circ$ & 2.5 & 3.23 &                        & 4$^\circ$ & 3.4 & 2.9  \\ %\hline
                       & -6$^\circ$ & 0.5 & 10.4 &                        & 6$^\circ$ & 3.5 & 4.0  \\ \hline
\end{tabular}
\label{tab:Beamsteering}
\end{table}

\subsection{Single Beam Control for Large Angles}
For the wave-controlled RIS with a single biasing frequency, significant side lobes are created which limit the effectiveness of single beam-steering when the required phase range exceeds $360^\circ$. However, extending the application of this single-frequency methodology to these larger angles could be useful in applications where lower directivity and large side lobes are acceptable. To study this case, we apply a method for optimizing $f_b$ and analyze the resulting radiation patterns.

To simplify the optimization of $f_\mathrm{b}$, the conditions under which the BSW's phase gradient approximates the slope of the ideal phase gradient are found. This occurs when $\lambda_\alpha = \lambda_\mathrm{b}/4$, where $\lambda_\alpha$ is the spatial period of the ideal wrapped phase gradient. The spatial period is defined as $\lambda_\alpha=|x_{m_1}-x_{m_2}|$, where $x_{m_{1,2}}$ are positions along the BTL with reflected phases $\alpha_{m_{1,2}}$ such that $|\alpha_{m_1}-\alpha_{m_2}|=2\pi$. Using (\ref{eq:SingleBeamElementPhase}), we find $\lambda_\alpha = |\sin{(\theta_\mathrm{p})|/\lambda_\mathrm{c}}$. This yields the following relationship between a given biasing frequency and the beam steering angle:
\begin{equation}\label{eq:DesignedAngle}
    f_\mathrm{b} = \frac{f_\mathrm{c} \left|\sin{(\theta_{\mathrm{p}})}\right|}{4 n_{\mathrm{slow}}}.
\end{equation}

\begin{figure}[t]
\label{Design}
\begin{centering}
\includegraphics[width=3.5in]{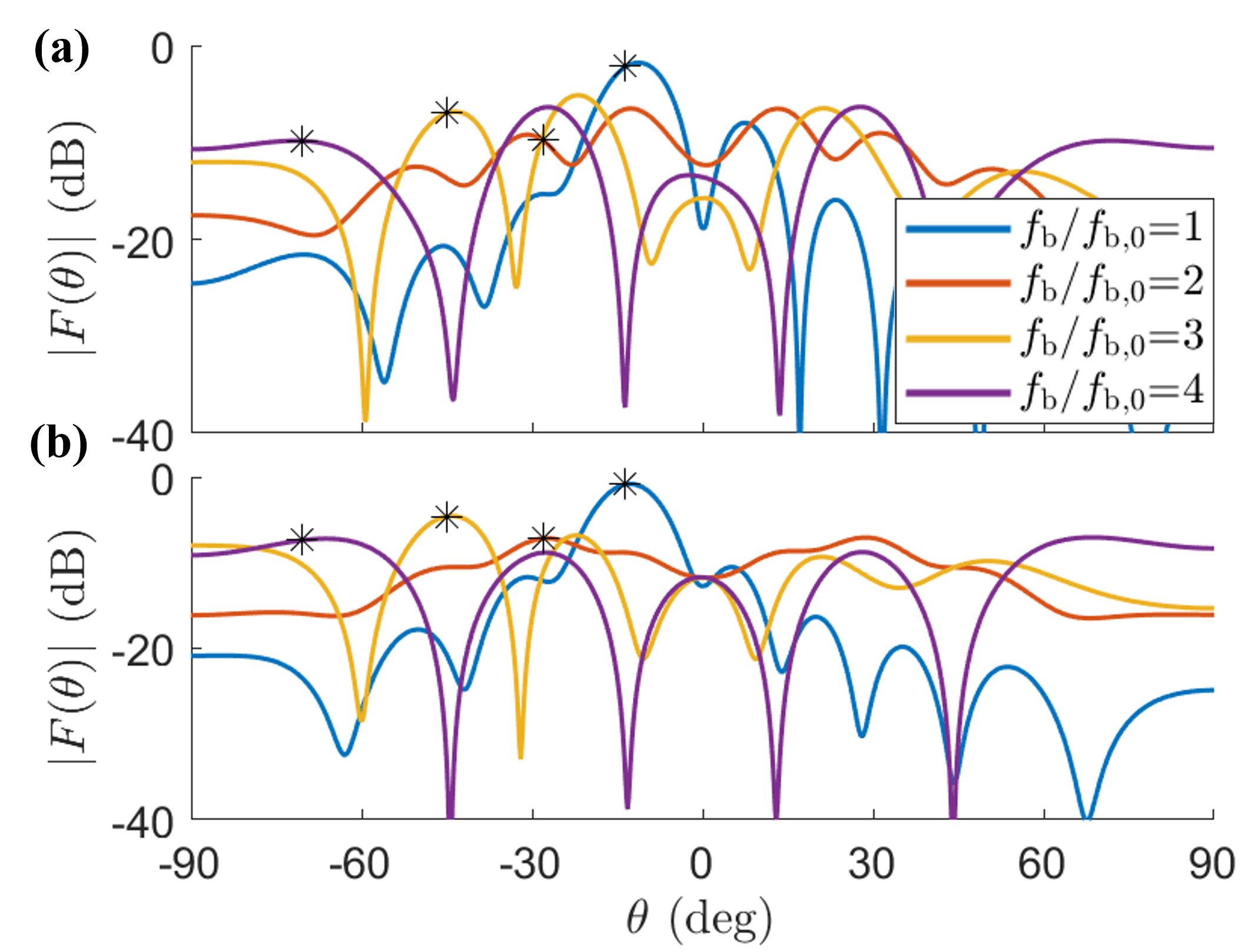}
\par\end{centering}
\caption{Normalized radiation patterns for the short-circuited wave-controlled RIS operating at $f_\mathrm{c} = 2.45\ \mathrm{GHz}$ for four biasing frequencies that maximize reflection at the marked angles found from (\ref{eq:DesignedAngle}). The $\theta_\mathrm{p}$ angles are $13.6^\circ, 28.1^\circ, 44.9^\circ,$ and $70.4^\circ$. The results are presented for two cases: (a) the designed RIS board; (b) an ideal and lossless RIS with elements that have a perfectly linear phase-vs-voltage relationship.
} \label{fig:optRadPattern}
\end{figure}

Several cases of radiation patterns using four biasing frequencies that are multiples of $f_{\mathrm{b},0}$ are shown in Fig.~\ref{fig:optRadPattern}, each leading to different peak-radiation angles. These radiation patterns illustrate the limitations of the technique discussed in this section. The blue line corresponds to the limit of the near-broadside beam steering when $|\alpha_{M-1}-\alpha_0|=360^\circ$. The orange and purple lines correspond to biasing frequencies that create symmetric standing wave voltages around the center of the BTL ($w(m) = w({M-1-m})$) in (\ref{eq:BTL_Volt_Retif_Single})) and thus symmetric beam-splitting radiation patterns around $\theta=0^\circ$. The line in yellow shows that, when the standing wave voltage is not symmetric and $|\alpha_{M-1}-\alpha_0|>360^\circ$, the radiation pattern has large sidelobes that reduce the directivity towards $\theta_\mathrm{p}$. 
This figure also compares the radiation patterns for the current design, shown in Fig.~\ref{fig:optRadPattern}(a), to those of a lossless design with unit cells that have a linear voltage-to-phase relationship for the carrier signal $f_\mathrm{c}$, shown in Fig.~\ref{fig:optRadPattern}(b). 
The linear relationship is calculated by $\alpha_m=360^\circ*(V-V_\mathrm{min}))/(V_\mathrm{max}-V_\mathrm{min})$ where $V$ is the voltage applied to the unit cell, and $V_\mathrm{min}=4$ and $V_\mathrm{max}=15$ are the respective minimum and maximum voltages allowed for the design.
%This illustrates, through the increase of reflection towards $\theta_\mathrm{p}$, 
These results illustrate that the beam steering method can be improved by using wider bandwidth elements \cite{Bradshaw_Bandwidth_25}.

An alternative approach for beam steering at large angles that uses the same optimization methods is to split the RIS into several segments with distinct BSWs. This method comes with certain tradeoffs \cite{Bradshaw_MultiBoard_2025}, but it avoids the creation of undesirable symmetric beams around $\theta = 0^\circ$. Additionally, the beam steering angle can be extended using multi-resonant unit cells with a phase variation limit larger than $360^\circ$ \cite{LinearElement}.

\section{Conclusions}

This paper has presented a comprehensive study of the design and modeling of a wave-controlled RIS, providing both analytical and numerical frameworks to characterize its performance. The designed RIS integrates a varactor-tuned unit cell along with a BTL and rectifiers that extract the required biasing voltage for the RIS. The effectiveness of this approach in generating the correct dc bias from a single standing wave frequency is validated through theoretical analysis and experimental verification. Furthermore, key design criteria for optimizing the waveguide voltages are identified, including the selection of an appropriate fundamental biasing frequency and its odd multiples, as well as matching the characteristic impedance of the BTL to the signal generator impedance, effectively doubling the number of biasing modes within the same frequency range. Additionally, this work demonstrates the feasibility of achieving reflective control of the RIS using a single standing wave frequency, paving the way for more efficient and cost-effective implementations of dynamically reconfigurable metasurfaces.

\section*{Acknowledgment}
This work is supported by the US National Science Foundation Award ECCS-2030029, and by a National Science Foundation scholarship funded under grant DUE-1930546. The authors thank DS SIMULIA for providing CST Studio Suite.

%\appendix  
%\section*{Appendix A\\Derivation of the Analytical Model of the RIS Unit Cell} \label{Appendix:A}  
\appendices 
\section{Derivation of the Analytical Model of the RIS Unit Cell} \label{Appendix:A}

The realistic RIS circuit model shown in Fig.~\ref{fig:Analytical_Model_RIS} is used to evaluate the reflection coefficient of the RIS \cite{Hanna22}, \cite{Costa21}. The equivalent impedance of the RIS, $Z_\mathrm{eq}$, seen at the surface level by a normally-incident plane wave, assuming that the varactor is not yet connected, is given by 
\begin{equation}\label{eq:zin}  
Z_\mathrm{eq}= \left (R_\mathrm{d}+j\omega L_\mathrm{d}+ \frac{1}{j\omega C_\mathrm{d}}\right )\, || \,j\omega L_\mathrm{s}.  
\end{equation}  
After some algebraic manipulations, this expression simplifies to  
%\begin{equation}  
%Z_\mathrm{eq}= \frac{j\omega L_\mathrm{s}\left (j\omega R_\mathrm{d}C_\mathrm{d}-\omega^{2}L_\mathrm{d}C_\mathrm{d}+1\right )}{j\omega C_\mathrm{d}\left (R_\mathrm{d}+j\omega\left (L_\mathrm{d} + L_\mathrm{s}\right )+1/\left ( j\omega C_\mathrm{d}\right )\right )},  
%\end{equation}  
%\noindent which can be further rewritten in fractional polynomial form as  
\begin{equation}  
Z_\mathrm{eq}= \frac{j\omega L_\mathrm{s}\left (1+j\omega R_\mathrm{d}C_\mathrm{d}-\omega^{2}\left (L_\mathrm{d}C_\mathrm{d}\right )\right )}{1+j\omega R_\mathrm{d}C_\mathrm{d}-\omega^{2}C_\mathrm{d}(L_\mathrm{d} + L_\mathrm{s})}.  
\end{equation}  

This expression can be reformulated in terms of the so-called electric resonance $\omega_\mathrm{e}$ and magnetic resonance $\omega_\mathrm{m}$. At $\omega_\mathrm{e}$, the reflection phase is nearly $180^\circ$, while when it is near (but not exactly at) $\omega_\mathrm{m}$, the reflection phase is approximately $0^\circ$. The magnetic resonance also plays a key role in manipulating the phase of the reflection coefficient and enabling its tunability \cite{Sievenpiper_HZ_EM_Surf_99, Best_Dipole_EBG_08, Capolino_Eq_TL_Model_13}. These two resonances are defined as  

\begin{equation}  
\omega_\mathrm{e}^2 = \frac{1}{C_\mathrm{d} L_\mathrm{d}}, \quad \omega_\mathrm{m}^2 = \frac{1}{C_\mathrm{d}\left (L_\mathrm{d}+L_\mathrm{s}\right )},
\end{equation}  

\noindent leading to the final expression for the impedance at the surface of the RIS unit cell:  

\begin{equation}  
Z_\mathrm{eq}= \frac{j\omega L_\mathrm{s}\left (1+j\omega R_\mathrm{d}C_\mathrm{d}-\left (\omega/\omega_\mathrm{e}\right )^{2}\right )}{1+j\omega R_\mathrm{d}C_\mathrm{d}-\left (\omega/\omega_\mathrm{m}\right )^{2}}.  
\label{eq:Zeq_we_wm}
\end{equation}  

The numerical values for the circuit elements $L_\mathrm{d}$, $C_\mathrm{d}$, and $R_\mathrm{d}$ are determined by performing a single full-wave simulation of the RIS unit cell shown in Fig. \ref{fig:RIS_Dimensions.png} without including the varactor, the biasing via at the center of the patch, or the coplanar waveguide on the RIS board ground plane (it includes the via to GND and the small rectangular patch connected to it that will connect to the varactor). In our case, we use CST Studio Suite to model the RIS under an orthogonally incident plane wave polarized along $y$ and traveling along the $-z$ direction, thereby accounting for mutual coupling among the RIS elements (we use periodic boundary conditions in $x$ and $y$). The model also incorporates copper losses and the relevant dielectric material properties. From this simulation, the scattering parameters are extracted, and the impedance $Z^\mathrm{FW}_\mathrm{eq}$ is computed from the Z-parameters to determine the values of $\omega_\mathrm{e}$ and $\omega_\mathrm{m}$.
For simplicity, $\omega_\mathrm{e}$ and  $\omega_\mathrm{m}$ are approximated by determining when the phase is exactly $180^\circ$ and $0^\circ$, respectively. Then, the circuit element values are obtained as follows:  
\begin{itemize}

\item The equivalent inductance $L_\mathrm{s}$ is analytically derived by modeling the dielectric substrate as a transmission line of length $D$ in the $z$ direction, with a ground plane termination. Using a Taylor series expansion, the impedance expression $j\omega L_{\mathrm{s}}=j\sqrt{\mu_0/(\epsilon_0 \epsilon_r)} \tan (\omega\sqrt{\epsilon_r}D /c)$ leads to the  approximate value  

\begin{equation}  
L_\mathrm{s}  \approx \mu_0D.  
\end{equation}  

\item Given $L_\mathrm{s}$, the inductance $L_\mathrm{d}$ is obtained as  

\begin{equation}  
L_\mathrm{d}= \frac{L_\mathrm{s}}{\left(\omega_\mathrm{e}/\omega_\mathrm{m}\right)^2-1}.  
\end{equation}  

\item With $L_\mathrm{d}$ known, the capacitance $C_\mathrm{d}$ is determined using  

\begin{equation}  
C_\mathrm{d} = \frac{1}{L_\mathrm{d} \omega_\mathrm{e}^2}.
\end{equation}  

\item Finally, the resistance $R_\mathrm{d}$ is obtained from the real part of $Z^\mathrm{FW}_\mathrm{eq}$ evaluated at $\omega_\mathrm{m}$ from the full-wave simulation (at this frequency, the imaginary part is independent of $R_\mathrm{d}$):

\begin{equation}  
R_\mathrm{d} = \frac{L_\mathrm{s}}{C_\mathrm{d}\left(1+L_\mathrm{d}/L_\mathrm{s}\right)\Re\left(Z^\mathrm{FW}_\mathrm{eq}(\omega_\mathrm{m})\right)}.  
\end{equation}  
\end{itemize}

%\appendix 
\section{Maximization of $W_\mathrm{b}$ - Short-circuit Termination}
\label{Appendix:B}
%$W_\mathrm{b} = \left|A \right|$

The amplitude $W_\mathrm{b}$ of the induced standing wave 
%due to the $n$th multiple of $f_\mathrm{b,0}$ 
can be calculated as a function of the voltage $V_\mathrm{in}$ at the connection of the BTL with the signal generator:
\begin{equation}
W_\mathrm{b} =\left|\frac{2V_\mathrm{in}}{e^{-j\kappa} - e^{j\kappa}} \right|, 
\end{equation}
where $\kappa= (\omega_\mathrm{b} n_{\mathrm{slow}}/c)L_{\mathrm{tot}}$. %$\kappa=k_\mathrm{b} L_{\mathrm{tot}}$
Assuming a BTL terminated in a short circuit, the input impedance of the BTL 
%for the $n$th multiple of $f_\mathrm{b,0}$ 
is $Z_\mathrm{in} = jZ_0 \mathrm{tan}\left( \kappa\right)$. Furthermore, assuming that at frequency $f_\mathrm{b}$ the capacitances $C_\mathrm{f}$ can be approximated as short circuits and the inductance $L_\mathrm{f}$ as an open circuit, we have  
\begin{equation}
V_\mathrm{in} = \frac{Z_\mathrm{in}}{Z_\mathrm{in} + Z_{\mathrm{g}}}V_{\mathrm{g}} = \frac{jZ_0\tan(\kappa)}{jZ_0\tan(\kappa) + Z_{\mathrm{g}}}V_{\mathrm{g}}.
\end{equation}
The expression for $W_\mathrm{b}$ can thus be written as a function of the impedances $Z_0$ and $Z_{\mathrm{g}}$:
\begin{equation}
W_\mathrm{b} =\left|\frac{2\frac{jZ_0\tan(\kappa)}{jZ_0\tan(\kappa) + Z_\mathrm{g}}V_\mathrm{g}}{e^{-j\kappa} - e^{j\kappa}} \right|.
\label{eq:W_n_definition}
\end{equation}

%From this expression, when $\kappa_n=\alpha\pi/2$, with $\alpha=1,3,5,\dots$, we have $\tan(\alpha\pi/2)\to \infty$ and $e^{-j\alpha\pi/2} - e^{j\alpha\pi/2}=-j2$, leading to $W_n=V_\mathrm{g}$, which is always at its maximum, as shown in Fig. \ref{fig:A_vs_SclFactor}. However, when $\kappa_n=\alpha\pi$, with $\alpha=1,2,3,\dots$, $\tan(\alpha\pi)=0$ while $e^{-j\alpha\pi} - e^{j\alpha\pi}\to0$, making it necessary to evaluate $W_n$ as $\kappa_n\to\alpha\pi$ to determine its dependency on the remaining parameters. %Such limit can be derived as

When considering the case with $f_\mathrm{b}=nf_\mathrm{b,0}$, the wavelength is $\lambda_\mathrm{b}=4L_\mathrm{tot}/n$, resulting in $\kappa=n\pi/2$. When $f_\mathrm{b}$ is an odd multiple of $f_\mathrm{b,0}$, we find that $\kappa=n \pi /2$, so $\tan(\kappa)=\infty$ and $e^{-j\kappa} - e^{j \kappa}=-j2$, leading to $W_\mathrm{b}=V_\mathrm{g}$, which as shown in Fig.~\ref{fig:A_vs_SclFactor} is always at its maximum value. However, when $f_\mathrm{b}$ is an even multiple of $f_\mathrm{b,0}$, $\tan(\kappa)=0$ while $e^{-j\kappa} - e^{j\kappa}=0$, making it necessary to evaluate $W_\mathrm{b}$ in the limit as $\kappa\to n\pi/2$ for even $n$, leading to
\begin{equation}
%\lim_{\kappa\to\alpha\pi} W_\mathrm{b} 
W_\mathrm{b} 
= \frac{Z_0}{Z_{\mathrm{g}}}V_{\mathrm{g}}.
\end{equation}

This result indicates that when $Z_0 = Z_{\mathrm{g}}$, the amplitude of the standing wave equals $V_{\mathrm{g}}$, the same value obtained when $f_\mathrm{b}$ is an odd multiple of $f_\mathrm{b,0}$. In fact, under this matching condition, the specific value of $\kappa$ becomes irrelevant. Indeed, evaluating (\ref{eq:W_n_definition}) with $Z_0 = Z_{\mathrm{g}}$ gives
\begin{equation}
W_\mathrm{b} \Bigr\rvert_{Z_0 = Z_{\mathrm{g}}} =\left|\frac{2\frac{j\tan(\kappa)}{j\tan(\kappa) + 1}V_\mathrm{g}}{e^{-j\kappa} - e^{j\kappa}} \right|.
\end{equation}
This expression simplifies to
\begin{align}
W_\mathrm{b} \Bigr\rvert_{Z_0 = Z_{\mathrm{g}}} &=\left|\frac{-\tan(\kappa)}{j\sin(\kappa)\tan(\kappa) + \sin(\kappa)}V_\mathrm{g} \right| = V_\mathrm{g}.
\end{align}

\bibliographystyle{IEEEtran}
\bibliography{Bibliography}

\begin{thebibliography}{10}
\providecommand{\url}[1]{#1}
\csname url@samestyle\endcsname
\providecommand{\newblock}{\relax}
\providecommand{\bibinfo}[2]{#2}
\providecommand{\BIBentrySTDinterwordspacing}{\spaceskip=0pt\relax}
\providecommand{\BIBentryALTinterwordstretchfactor}{4}
\providecommand{\BIBentryALTinterwordspacing}{\spaceskip=\fontdimen2\font plus
\BIBentryALTinterwordstretchfactor\fontdimen3\font minus \fontdimen4\font\relax}
\providecommand{\BIBforeignlanguage}[2]{{%
\expandafter\ifx\csname l@#1\endcsname\relax
\typeout{** WARNING: IEEEtran.bst: No hyphenation pattern has been}%
\typeout{** loaded for the language `#1'. Using the pattern for}%
\typeout{** the default language instead.}%
\else
\language=\csname l@#1\endcsname
\fi
#2}}
\providecommand{\BIBdecl}{\relax}
\BIBdecl

\bibitem{Sharma_Reconfigurable_21}
T.~Sharma, A.~Chehri, and P.~Fortier, ``Reconfigurable intelligent surfaces for {5G} and beyond wireless communications: A comprehensive survey,'' \emph{Energies}, vol.~14, no.~24, 2021.

\bibitem{Abeywickrama20}
S.~Abeywickrama, R.~Zhang, Q.~Wu, and C.~Yuen, ``Intelligent reflecting surface: Practical phase shift model and beamforming optimization,'' \emph{IEEE Transactions on Communications}, vol.~68, no.~9, pp. 5849--5863, 2020.

\bibitem{Costa21}
F.~Costa and M.~Borgese, ``Electromagnetic model of reflective intelligent surfaces,'' \emph{IEEE Open Journal of the Communications Society}, vol.~2, pp. 1577--1589, 2021.

\bibitem{Yang_Beyond_24}
H.~Yang, S.~Kim, H.~Kim, S.~Bang, Y.~Kim, S.~Kim, K.~Park, D.~Kwon, and J.~Oh, ``Beyond limitations of {5G} with {RIS}: Field trial in a commercial network, recent advances, and future directions,'' \emph{IEEE Communications Magazine}, vol.~62, no.~10, pp. 132--138, 2024.

\bibitem{Du_Millimeter_21}
H.~Du, J.~Zhang, J.~Cheng, and B.~Ai, ``Millimeter wave communications with reconfigurable intelligent surfaces: Performance analysis and optimization,'' \emph{IEEE Transactions on Communications}, vol.~69, no.~4, pp. 2752--2768, 2021.

\bibitem{cui_coding_14}
T.~J. Cui, M.~Q. Qi, X.~Wan, J.~Zhao, and Q.~Cheng, ``Coding metamaterials, digital metamaterials and programmable metamaterials,'' \emph{Light: Science \& Applications}, vol.~3, no.~10, pp. e218--e218, 2014.

\bibitem{Basar_Wireless_19}
E.~Basar, M.~Di~Renzo, J.~De~Rosny, M.~Debbah, M.-S. Alouini, and R.~Zhang, ``Wireless communications through reconfigurable intelligent surfaces,'' \emph{IEEE Access}, vol.~7, pp. 116\,753--116\,773, 2019.

\bibitem{Hershko_Photodiode_24}
T.~Hershko, D.~Rozban, G.~Kedar, A.~Etinger, and A.~Abramovich, ``Photodiode as tuning element for metasurface intelligent reflecting surface based on steer by image technology,'' in \emph{Proc. IEEE International Symposium on Antennas and Propagation and INC/USNC‐URSI Radio Science Meeting}, 2024, pp. 825--826.

\bibitem{Sayanskiy_2D-Programmable_23}
A.~Sayanskiy, A.~Belov, R.~Yafasov, A.~Lyulyakin, A.~Sherstobitov, S.~Glybovski, and V.~Lyashev, ``A {2D}-programmable and scalable reconfigurable intelligent surface remotely controlled via digital infrared code,'' \emph{IEEE Transactions on Antennas and Propagation}, vol.~71, no.~1, pp. 570--580, 2023.

\bibitem{Miao_Light-Controlled_23}
S.~Y. Miao and F.~H. Lin, ``Light-controlled large-scale wirelessly reconfigurable microstrip reflectarrays,'' \emph{IEEE Transactions on Antennas and Propagation}, vol.~71, no.~2, pp. 1613--1622, 2023.

\bibitem{Hu_Ultrafast_19}
Y.~H. et~al, ``Ultrafast terahertz frequency and phase tuning by all-optical molecularization of metasurfaces,'' \emph{Advanced Optical Materials}, vol.~7, no.~22, p. 1901050, 2019.

\bibitem{Zhang_Light-Controllable_18}
\BIBentryALTinterwordspacing
X.~G. Zhang, W.~X. Tang, W.~X. Jiang, G.~D. Bai, J.~Tang, L.~Bai, C.-W. Qiu, and T.~J. Cui, ``Light-controllable digital coding metasurfaces,'' \emph{Advanced Science}, vol.~5, no.~11, p. 1801028, 2018. [Online]. Available: \url{https://advanced.onlinelibrary.wiley.com/doi/abs/10.1002/advs.201801028}
\BIBentrySTDinterwordspacing

\bibitem{Ayanoglu22}
E.~Ayanoglu, F.~Capolino, and A.~L. Swindlehurst, ``Wave-controlled metasurface-based reconfigurable intelligent surfaces,'' \emph{IEEE Wireless Communications}, vol.~29, no.~4, pp. 86--92, 2022.

\bibitem{Saavedra-Melo_APS_RIS_23}
M.~Saavedra-Melo, K.~Rouhi, and F.~Capolino, ``Wave-controlled {RIS}: A novel method for reconfigurable elements biasing,'' in \emph{Proc. IEEE International Symposium on Antennas and Propagation and USNC-URSI Radio Science Meeting (USNC-URSI)}, 2023, pp. 979--980.

\bibitem{Saavedra-Melo_APS_RIS_24}
M.~Saavedra-Melo, V.~Yao, B.~Bradshaw, K.~Rouhi, A.~De~Leon, A.~Nikzamir, R.~Marosi, and F.~Capolino, ``Progress in wave-controlled {RIS}: Experimental demonstration of biasing voltage generation,'' in \emph{Proc. IEEE International Symposium on Antennas and Propagation and INC/USNC‐URSI Radio Science Meeting (AP-S/INC-USNC-URSI)}, 2024, pp. 1963--1964.

\bibitem{Ben-Itzhak24}
G.~Ben-Itzhak, M.~Saavedra-Melo, B.~Bradshaw, E.~Ayanoglu, F.~Capolino, and A.~Lee~Swindlehurst, ``Design and operation principles of a wave-controlled reconfigurable intelligent surface,'' \emph{IEEE Open Journal of the Communications Society}, vol.~5, pp. 7730--7751, 2024.

\bibitem{Saavedra-Melo_LACAP_RIS_24}
M.~Saavedra-Melo, B.~Bradshaw, and F.~Capolino, ``Analysis of the experimental bias voltage generation for {RIS} using standing wave control,'' in \emph{Proc. 1st IEEE Latin American Conference on Antennas and Propagation (LACAP)}, 2024.

\bibitem{Ben-Itzhak2025AI-Driven}
G.~Ben-Itzhak, M.~Saavedra-Melo, E.~Ayanoglu, F.~Capolino, and A.~L. Swindlehurst, ``{AI}-driven optimization of wave-controlled reconfigurable intelligent surfaces,'' \emph{IEEE Open Journal of the Communications Society}, vol.~6, pp. 6650--6665, 2025.

\bibitem{Hanna22}
D.~Hanna, M.~Saavedra-Melo, F.~Shan, and F.~Capolino, ``A versatile polynomial model for reflection by a reflective intelligent surface with varactors,'' in \emph{Proc. IEEE International Symposium on Antennas and Propagation and USNC-URSI Radio Science Meeting (AP-S/URSI)}, 2022, pp. 679--680.

\bibitem{pozar_microwave_2012_ch2}
D.~M. Pozar, ``Transmission line theory,'' in \emph{Microwave Engineering}, 4th~ed.\hskip 1em plus 0.5em minus 0.4em\relax Hoboken, NJ, USA: Wiley, 2012, ch.~2, pp. 48--89.

\bibitem{ZhangIEEETAP2003PlanarAMC}
Y.~Zhang, J.~von Hagen, M.~Younis, C.~Fischer, and W.~Wiesbeck, ``Planar artificial magnetic conductors and patch antennas,'' \emph{IEEE Transactions on Antennas and Propagation}, vol.~51, no.~10, pp. 2704--2712, 2003.

\bibitem{donzelli2009metamaterial}
G.~Donzelli, A.~Vallecchi, F.~Capolino, and A.~Schuchinsky, ``Metamaterial made of paired planar conductors: Particle resonances, phenomena and properties,'' \emph{Metamaterials}, vol.~3, no.~1, pp. 10--27, 2009.

\bibitem{pozar_microwave_2012_ch3}
D.~M. Pozar, ``Transmission lines and waveguides,'' in \emph{Microwave Engineering}, 4th~ed.\hskip 1em plus 0.5em minus 0.4em\relax Hoboken, NJ, USA: Wiley, 2012, ch.~3, pp. 148--149.

\bibitem{Cheng_RCS_Metasurf_15}
C.~Huang, W.~Pan, X.~Ma, and X.~Luo, ``Wideband radar cross-section reduction of a stacked patch array antenna using metasurface,'' \emph{IEEE Antennas and Wireless Propagation Letters}, vol.~14, pp. 1369--1372, 2015.

\bibitem{Xu_Radar_Dyn_Jamming_21}
H.~Xu, D.-F. Guan, B.~Peng, Z.~Liu, S.~Yong, and Y.~Liu, ``Radar one-dimensional range profile dynamic jamming based on programmable metasurface,'' \emph{IEEE Antennas and Wireless Propagation Letters}, vol.~20, no.~10, pp. 1883--1887, 2021.

\bibitem{Bradshaw_MultiBoard_2025}
B.~Bradshaw, M.~Saavedra-Melo, and F.~Capolino, ``Wave-controlled {RIS}: Biasing {RIS} elements using standing waves with single frequency,'' in \emph{Proc. 19th European Conference on Antennas and Propagation (EuCAP)}, 2025.

\bibitem{Balanis_Antenna_15}
C.~A. Balanis, \emph{Antenna Theory: Analysis and Design}.\hskip 1em plus 0.5em minus 0.4em\relax John Wiley \& Sons, 2015.

\bibitem{LinearElement}
Q.~Hu, H.~Yang, X.~Zeng, and X.~Y. Zhang, ``Wideband reconfigurable intelligent surface using dual-resonance element,'' \emph{IEEE Antennas and Wireless Propagation Letters}, vol.~22, no.~10, pp. 2422--2426, 2023.

\bibitem{Bradshaw_Bandwidth_25}
B.~Bradshaw, M.~Saavedra-Melo, S.~Sharma, and F.~Capolino, ``On {RIS} bandwidth: {B}andwidth discussion and trade-offs for a dual resonant, dual linearly polarized element,'' in \emph{Proc. U.S. National Committee of URSI National Radio Science Meeting (USNC-URSI NRSM)}, 2025, pp. 383--384.

\bibitem{Sievenpiper_HZ_EM_Surf_99}
D.~Sievenpiper, L.~Zhang, R.~Broas, N.~Alexopolous, and E.~Yablonovitch, ``High-impedance electromagnetic surfaces with a forbidden frequency band,'' \emph{IEEE Trans. on Microwave Theory and Techniques}, vol.~47, no.~11, pp. 2059--2074, 1999.

\bibitem{Best_Dipole_EBG_08}
S.~R. Best and D.~L. Hanna, ``Design of a broadband dipole in close proximity to an {EBG} ground plane,'' \emph{IEEE Antennas and Propagation Magazine}, vol.~50, no.~6, pp. 52--64, 2008.

\bibitem{Capolino_Eq_TL_Model_13}
F.~Capolino, A.~Vallecchi, and M.~Albani, ``Equivalent transmission line model with a lumped {X}-circuit for a metalayer made of pairs of planar conductors,'' \emph{IEEE Transactions on Antennas and Propagation}, vol.~61, no.~2, pp. 852--861, 2013.

\end{thebibliography}


\begin{thebibliography}{1}
\bibitem{Costa21}
F. Costa and M. Borgese, ``Electromagnetic Model of Reflective Intelligent Surfaces,'' in \emph  , 2021, doi: 10.1109/OJCOMS.2021.3092217.
\bibitem{Ayanoglu22}
E. Ayanoglu, F. Capolino and A. L. Swindlehurst, ``Wave-Controlled Metasurface-Based Reconfigurable Intelligent Surfaces,'' in \emph{IEEE Wireless Communications}, vol. 29, no. 4, pp. 86-92, August 2022, doi: 10.1109/MWC.005.2100401.
%\bibitem{Hanna22}
%D. Hanna, M. S. Melo, F. Shan and F. Capolino, ``A Versatile Polynomial Model for Reflection by a Reflective Intelligent Surface with Varactors,'' \emph{2022 IEEE International Symposium on Antennas and Propagation and USNC-URSI Radio Science Meeting (AP-S/URSI)}, Denver, CO, USA, 2022, pp. 679-680, doi: 10.1109/AP-S/USNC-URSI47032.2022.9887248.
\bibitem{Saavedra-Melo23}
M. Saavedra-Melo, K. Rouhi and F. Capolino, ``Wave-Controlled RIS: A Novel Method for Reconfigurable Elements Biasing,'' \emph{2023 IEEE International Symposium on Antennas and Propagation and USNC-URSI Radio Science Meeting (USNC-URSI)}, Portland, OR, USA, 2023, pp. 979-980, doi: 10.1109/USNC-URSI52151.2023.10237459.

\end{thebibliography}

\begin{comment}
    
\end{comment}

\end{document}